\documentclass[preprint,prb,superscriptaddress,aps]{revtex4}
\usepackage{graphicx}
\usepackage{float}
\usepackage{dcolumn}
\usepackage{amsmath,amssymb}
\usepackage{multirow}
\usepackage{boxedminipage}
\usepackage{array}
\usepackage{flafter}
\usepackage[mathscr]{euscript}
\usepackage[tight,TABTOPCAP]{subfigure}
\usepackage[normalsize]{caption}
\usepackage{verbatim}
\usepackage{xcolor}
\usepackage[normalem]{ulem}
\newcolumntype{.}{D{.}{.}{1}}
\usepackage{bm}
\usepackage[utf8]{inputenc}
\usepackage[T1]{fontenc}
\usepackage{mathptmx}
\usepackage{etoolbox}
\makeatletter
\def\@email#1#2{%
 \endgroup
 \patchcmd{\titleblock@produce}
  {\frontmatter@RRAPformat}
  {\frontmatter@RRAPformat{\produce@RRAP{*#1\href{mailto:#2}{#2}}}\frontmatter@RRAPformat}
  {}{}
}%
\makeatother

\begin{document}

\title{
A kinetic model of electron transfer at the electrode-electrolyte interface: Statistical mechanics and electrochemical
aspects
}

\author{Diego Veloza-Diaz}
\affiliation{Institut f{\"u}r Physik, Johannes Gutenberg-Universit{\"a}t Mainz, Staudingerweg 9, 55128 Mainz, Germany}
\author{Robinson Cortes-Huerto$^{*}$}
\affiliation{Max Planck Institute for Polymer Research, Ackermannweg 10, 55128, Mainz, Germany}
\affiliation{Present address: DWI – Leibniz Institute for Interactive Materials, \\
52074, Aachen, Germany}
\email{corteshu@mpip-mainz.mpg.de.}
\author{Pietro Ballone}
\affiliation{Max Planck Institute for Polymer Research, Ackermannweg 10, 55128, Mainz, Germany}
\author{Nancy C. Forero-Martinez$^{*}$}
\affiliation{Max Planck Institute for Polymer Research, Ackermannweg 10, 55128, Mainz, Germany}
\email{nforerom@uni-mainz.de.}

\date{\today}
\begin{abstract} A new simulation approach ({\it J. Chem. Phys.} {\bf 163}, 164113), based on the seamless combination of the 
electron transfer at the electrode/electrolyte interface with the charge and mass transport in electrolyte solutions, holds the 
promise to advance the comprehensive modeling of basic electrochemical devices such as electrolytic cells and batteries. The 
underlying model also includes a fully consistent treatment of structural, thermodynamic and statistical properties of the 
electrical double-layer at non-ideal electron-conducting interfaces. As a result, subtle aspects such as Tafel and Butler-Volmer
equations, the overpotential concept, as well as (stationary) non-equilibrium features such as the entropy production, arise 
naturally from the model, with a minimum of ad-hoc assumptions and elaborations of the simulation results. We illustrate these 
aspects for a simple model of electrochemical interface, and discuss further improvements of the method meant to enhance its 
ability to model systems and phenomena of interest for electrochemistry. One such improvement, concerning the effect of 
fluctuating electric fields on the electron transfer rate, is explicitly developed.
\end{abstract}

\maketitle

\section{Introduction}
\label{intro}
Electrochemistry arguably represents both a precursor and a current pillar of modern science,\cite{ecs} notable for its long 
history and for its ability to connect topics traditionally considered as separate, such as, for instance, the kinetics of 
chemical reactions, the electronic structure of metals, the statistical mechanics of fluid and solid electrolytes including, in 
particular, their mass and charge transport, the structural and thermodynamic properties of electrified interfaces.\cite{bock1,
bock2A, bock2B} This conceptual interest of electrochemistry is paralleled and often overshadowed by its role in a myriad of 
industrial and engineering applications,\cite{applied} in fields such as power generation and energy storage, chemical 
transformations driven by electric currents, metal extraction and purification, electroplating, electro-forming, etc. In this 
context, batteries and electrolytic cells represent paradigmatic devices, with fuel cells rapidly gaining ground with respect to
other power generation approaches.\cite{ener}

The non-equilibrium character of the stationary processes of interest, the role of interfaces with their lack of isotropy and 
low overall symmetry, the interplay of quantum and classical dynamical aspects, as well as the broad range of size and time 
scales encompassed by electrochemical processes pose a formidable challenge to any advance in the theoretical/computational 
exploration of electrochemistry. The difficulty has been compounded with the fact that the electrochemistry community, although
largely overlapping with the statistical mechanics and chemical-physics communities, has developed a somewhat specialised 
language, driven by its long history and specific (manly experimental) methods. 

Until now, the whole subject has been attacked primarily along an analytical route, i.e., focusing in turn on a sequence of 
separate sub-problems. In the early stages, for instance, the properties of idealised models of electrode/electrolyte 
interfaces have been analysed using traditional statistical mechanics methods (Helmholtz,\cite{hel} 
Gouy-Chapman,\cite{gouy, chap} Stern\cite{stern} models). Some time later, equilibrium computational methods and computer 
simulation in particular,\cite{scalfi} provided crucial insight on the same subject of electrified interfaces, starting from the
pioneering studies of Ref.~\onlinecite{valleau, torrie}. In a following stage, advances in the understanding of the electronic 
structure of metal surfaces\cite{newns1, newns2, schmic0} and electrified interfaces,\cite{norsk, ross} provided a way to 
interpret a number of experimental observation, as well as a computational framework for the systematic ab-initio (primarily 
density functional) determination of the electron transfer rate\cite{schmic2022, fang} at metal (and semiconducting) interfaces 
with electrolyte solutions. Also, the effect of the solution dynamics on the transfer rate was modelled and investigated by 
stochastic molecular dynamics.\cite{schmic2002} As an auxiliary effort, the connection with quantum chemistry methods was 
spelled out in detail,\cite{qm} offering an additional path for the step-by-step interpretation of the electron transfer 
process.

Early applications of ab-initio molecular dynamics (ab-initio MD) methods promised to transform and energize the entire 
field,\cite{price, lozovoi, pasquarello} moving towards the comprehensive and simultaneous analysis of electronic and ionic 
transport mechanisms. However, the rigorous matching of the ionic and electronic time evolution is hampered by the 
approximations required to make ab-initio MD a computationally viable algorithm. Keeping the discussion short and simple, we 
observe that, almost without exceptions, ab-initio MD methods describe the electronic structure in terms of density functional 
theory, often in the Kohn-Sham (KS) picture, in which the electron density is expressed through auxiliary KS orbitals, whose 
identification with physical single electrons may be deceptive. Second, and more importantly, the time evolution of the orbitals
is not the result of the application of a suitable unitary operator, as normally dictated by quantum mechanics. Instead, again 
almost without exceptions, it conforms to the Born-Oppenheimer approximation, in which the atomic nuclei move according to the
forces determined by the electron density, while the electrons remain in their ground state corresponding to the instantaneous 
configuration of the atoms. In this way, while the dynamics of the electrode atoms and electrolyte ions might be accurate, the 
flow of electrons across the electrode/electrolyte interface deviates from their true dynamics. At the same time, the electron 
population loses its particle-like character, hence the dynamics of discrete electrons is replaced by that of a continuous 
permeating fluid, i.e., the electron density. Perhaps more importantly, the stochastic nature of the electrons' motion is lost, 
replaced by a fully deterministic evolution. Whenever this evolution follows the adiabatic approximation, its characteristic 
times are determined by the motion of the atoms and ions, disconnected from the energy and spatial distribution of the 
electronic levels which determine the true quantum mechanical electron dynamics. Besides these drawbacks of principle, one 
should also consider that, despite the breathtaking progress of computer facilities and computational methods of the last forty 
years, ab-initio simulations are still time consuming and expensive, limiting the range of sizes and especially times that can 
be covered by these methods.

Possibly for these reasons, the ab-initio and other microscopic/atomistic MD methods do not appear to be as closely integrated 
in mainstream electrochemistry as desirable,\cite{limits} certainly playing a role in elucidating structural and dynamical 
features of metal/electrolyte interfaces, but far from having realized their full potential in the description of 
electrochemical systems in all their essential aspects, which include charge transfer across interfaces and steady state flow in
each distinct (electrode and electrolyte) phase.

It might be worth noticing that the computational modeling of charge flowing trough purely electronic (semiconducting) devices 
has been developed before any ab-initio MD, and, at the middle of the '80s of last century,\cite{jaco} it was already a well 
established research tool, able to greatly contribute to the explosive development of micro-electronics of the last forty years.
This observation confirms that the major difficulty in the electrochemical case is the smooth and consistent matching of the 
ions and electrons transport properties, resulting from classical and quantum mechanical dynamics, respectively. A related 
observation is that the approaches developed to investigate the flow of electrons in semiconducting devices relied almost 
exclusively on stochastic methods to evolve the electron population in time, simultaneously reflecting the quantum nature of the
electron propagation, and enjoying a great efficiency advantage with respect to the deterministic MD algorithms used to simulate
the charge structure and dynamics in electrochemical systems.

Recently, a simple kinetic model\cite{eletransf} was proposed to simulate the transfer of electrons at the electrode/electrolyte
interface, coupled to the drift of ionic charge due to the gradient of the electrochemical potential in the electrolyte. For the
sake of clarity and generality, the approach was presented in minimalistic form, applied to idealised models of electrode, 
electrolyte and electron transfer kinetics. Explicit computations, using Monte Carlo (MC) to simulate the kinetics of the 
coupled electron transfer and ionic charge mobility, have shown that the model is able to reproduce the steady state flow of 
charge throughout this idealised electrochemical half-cell. Computational experiments, then, can be carried out to determine how
the current density depends on a variety of parameters and conditions, such as the surface charge density at the electrode, the 
ionic strength of the electrolyte, the diffusion constant of the ions, and the intrinsic rate of electron transfer from(/to) the
metal to(/from) the ions. Despite the many intriguing features, the scheme discussed in Ref.~\onlinecite{eletransf} is 
admittedly still very idealised and incomplete. 

The first aim of the present contribution is to highlight that, even in the original idealised version, the model offers a view
on properties and concepts such as Tafel and Butler-Volmer equations, overpotential, irreversibility and entropy generation that
are inherently intertwined with electrochemistry,\cite{bock2A} but whose determination by traditional approaches or even by 
ab-initio simulation is not always direct and intuitive. Their transparent determination by the kinetic approach will ease the
investigation of these quantities and of the relations among them, offering a more intimate contact with the electrochemical 
literature.

A second aim of the present study is to show how aspects neglected in the original publication\cite{eletransf} can be 
incorporated, extending the reach of the method and increasing its interest and realism. Feedback mechanisms that enhance or
hinder the flow through the electrode interface are important factors introducing non-linear relations among the properties of 
electrochemical systems. For instance, the application of an overpotential required to drive a non-equilibrium current through 
the device will result in the presence of sizeable fluctuating electric fields at the electrode surface that, in turn, affect 
the overall charge transfer rate at the interface and change also the time correlation among single electron transfer events.
It is shown that this feedback mechanism can be incorporated exactly and at a modest computational cost.

Moreover, we take the opportunity to improve our notation, to distinguish quantities that were lumped together in the previous 
paper, and to make contact whenever possible with popular chemical-physics textbook expositions of the same topics.\cite{atk}

\begin{figure}[!htb]
\vskip 0.7truecm
\includegraphics[scale=0.45,angle=-0]{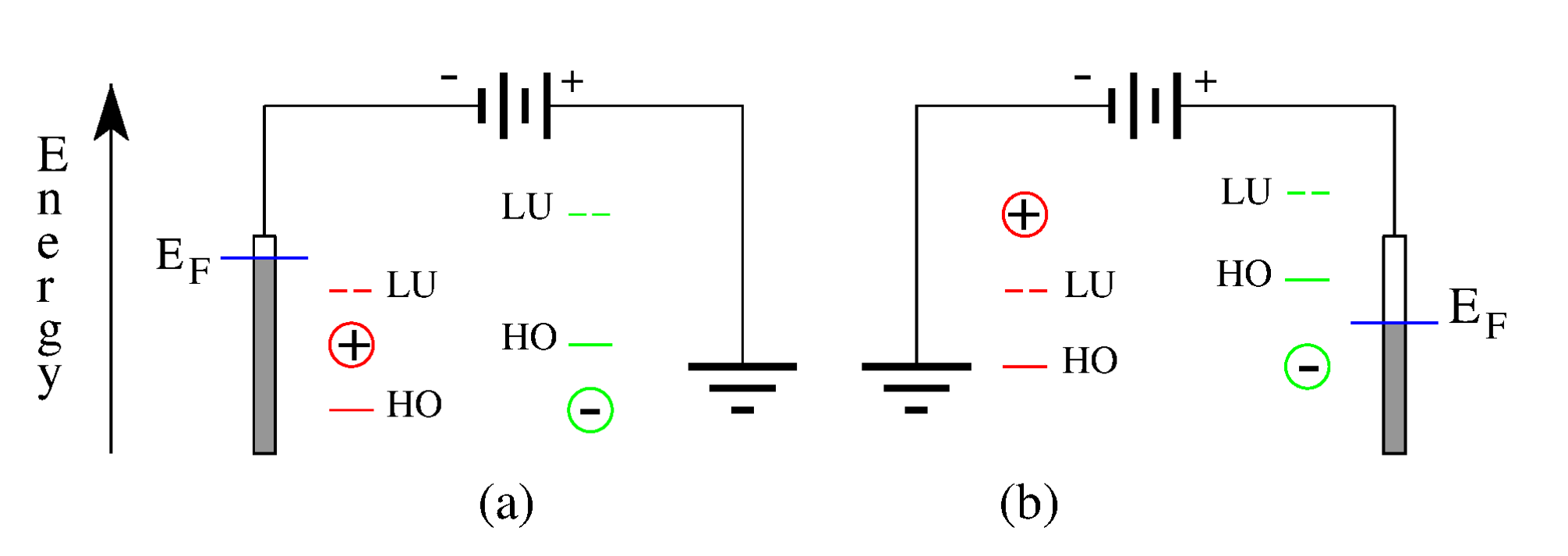}
\vskip 0.3truecm
\caption{Schematic diagram of the half-cell model. (a) Reduction of cations at the cathode; (b) oxidation of anions at the 
anode. The electrons flow from right to left, the electric current flows from left to right. The electron energy is 
measured on the common scale indicated by the vertical axis on the left. $E_F$ is the Fermi energy of the electrons in each
system. The atomic-like energy levels on the ions are indicated by a continuous (HOMO (H): highest occupied molecular
orbital) and dashed (LUMO (L): lowest unoccupied molecular orbital) segment, respectively. Weak hybridization with the metal
electron bands of the electrode turn each of these $\delta$-like states into a narrow but regular distribution.
}
\label{elelev}
\end{figure}

\section{The model and method}
\label{method}
\subsection{The model}
\label{mode}
The kinetic approach of Ref.~\onlinecite{eletransf} was developed on top of an ideally polarisable electrode/electrolyte 
interface model, whose electrode consisted of a rigid and structureless surface, while the electrolyte was represented by the 
primitive model, an implicit solvent model of a salt dissolved and fully dissociated in a liquid medium. A uniform electronic
surface charge density $\sigma_e$ resides on the electrode surface, and cannot cross the interface to react with the electrolyte
ions. In a similar way, ions from the electrolyte do not change their oxidation state and cannot penetrate the 
electrode surface beyond a position of closest approach to be specified below.\cite{valleau, torrie}

In the primitive model of electrolyte, the ion-ion interaction is of the hard-sphere plus Coulomb type:
\begin{equation}
v(\mid {\bf r_i-r_j}\mid)=\left\{
\begin{tabular}{ll}
$+\infty$                                    &\ \ \  $r_{ij} \leq d_{ij}$ \\ &                     \\
$ \frac{q_iq_j}{\epsilon r_{ij}}$ &\ \ \  $r_{ij} > d_{ij}$ \\
\end{tabular}
\right.
\label{hamilt1}
\end{equation}
where  $r_{ij} = \mid {\bf r_j-r_i}\mid$ is the distance of ions $i$ and $j$ and $d_{ij}$ is the distance of closest approach
of ions $i$ and $j$, for which additivity is assumed:
\begin{equation}
d_{ij}=\frac{(d_{ii}+d_{jj})}{2}
\end{equation}
In the examples discussed below, the solvent is water, which enters the model definition through the static dielectric constant 
(relative permittivity) $\epsilon=78.5$ in Eq.~\ref{hamilt1}. In what follows, charges are expressed in units of the (positive) 
electron charge $e$, while distances will be expressed in units of the cation diameter $d_{++}=1$. To make contact with real 
electrochemical systems, in the computations described in the following section, it is assumed that $q_i=\pm e$, $d_{++}=4.25$ 
\AA\ , $T=298$ K, and the scaled interaction among ions $\beta^{\ast}=e^2/(\epsilon d_{++} k_B T)$ is equal to $1.6809$. Time is
measured in nanoseconds (ns). This last choice is arbitrary, but it its usage is consistent with the prescription used to 
transform the MC time into real time, matching experimental diffusion constants of the ions, as briefly stated below. All other 
quantities are expressed in units derived from these basic ones. Energy, for instance, is expressed in $(e^2/d_{++})$ 
units, the electrostatic potential in $(e/d_{++})$, the electric current density in [$e/(d_{++}^2 ns)$], the resistivity
per unit area ${\cal{R}}$ in ($ns/d_{++}$), etc. 

The simulations discussed below use the peculiar geometry of the interface introduced in Ref.~\onlinecite{eletransf}. The 
electrolyte, in particular, occupies the cavity enclosed within the spherical surface of radius $R_e$ representing the 
electrode. Together with the grand-canonical MC (GC-MC) simulation method to fix the state of the electrolyte, the model 
provides a single-electrode approach to investigate electrified interfaces free of undue interactions with the complementary 
interface that completes any real electrochemical device. 

The idealised model of polarisable electrified interface is turned into a model of electrochemical half-cell by introducing a 
mechanism to transfer electrons across the electrode/electrolyte interface, which, in this way, loses its ideally polarisable 
character. Implicitly, this assumes an ordering of the relative energy levels of electrons on the electrode and on electrolyte 
ions which is schematically shown in Fig.~\ref{elelev}. The schemes in panel (a) and (b), in particular, are both drawn as being
part of an electrochemical cell, but can also be adapted to represent Galvanic half-cells,from which batteries can be made. A 
look at panel (a), for instance, shows that electrons at the Fermi level spontaneously reduce cations in the electrolyte 
solution. In panel (b), instead, electrons from anions cross the boundary with the electrode, forcing a charge to travel along 
the wire to reach their ground state.

The charge transfer process is primarily quantum mechanical, with, however, a dependence, possibly important, on classical
degrees of freedom. The first aim of any theory of this process is to compute the rate $k_e$ of electron transfer between the 
electrode and a redox centre in its proximity.\cite{schmic0} The rate will depend on their geometrical separation (and relative 
orientation, in the case of molecules), the oxidation state of the redox centre and the relative energy of its localised states 
with respect to the electrode Fermi energy. These aspects can be accounted for by a simple Hamiltonian acting on delocalised 
states from the metal electrode and localised states for the redox centre, with an interaction term to connect them. To obtain 
a model at least qualitatively correct, however, a further component has to be included, describing a bath of harmonic 
oscillators representing the solvent.\cite{schmic0, schmic2002} Modes coupled to the electron transfer, in particular, will 
account for the reorganisation energy which is a peculiar feature in Marcus and Hush theory of electron transfer.\cite{mushI, 
mushII, mushIII} It might be useful to remark that the quantum mechanical matrix elements underlying the determination of $k_e$ 
are the same for the direct reaction (say, M$^+$ + e$^-$ $\leftrightharpoons$ M) and the reciprocal one: M $\leftrightharpoons$ 
M$^+$ + e$^-$. These quantum mechanical matrix elements, in particular, connect initial and final states of the same 
energy,\cite{flet0, qm} their balance being achieved via fluctuations in the ionic atmosphere of electrolyte 
ions,\cite{fletI, fletII} or through the reorientation of solvent dipoles.\cite{mushI}

The inherent rate $k_e$, measured in electron-transfer events per ion and per unit time, should not be confused with the
overall reaction rates $h_o$ or $h_r$ (oxidation, reduction, respectively) at the interface, measured in electron-transfer 
events per unit area and unit time. These overall rates depend primarily on the number and distribution of
electrolyte ions in proximity of the interface, but may also depend on geometric (planar, curved) and structural (smooth or 
rough, stepped, defective, etc.) properties of the metal electrode surface. Finally, the radial electric current density 
$j_r(R_e)$ through the interface results from the combination (with the appropriate sign) of the overall rate of all competing 
or complementary reactions taking place at the same (cathode, anode) interface, and may be affected by the electron current 
flowing through the conducting wire, driven by the reactions at the reciprocal (anode, cathode) interface.
 
A crucial property of the approach discussed in Ref.~\onlinecite{eletransf} is to match the electron transfer part with the 
ionic conduction part, putting these first-order kinetic processes on a unique time scale. As explained in 
Ref.~\onlinecite{eletransf}, the self-diffusion coefficient of ions is used as the primary clock that marks the flow of the 
Monte Carlo time (measured in Monte Carlo attempted moves) into real time units, such, as, for instance, ns. Monte Carlo in a 
few of its variants (canonical, grand-canonical, kinetic) is used to evolve the system in time in a way that reflects the 
relative frequency of the various processes that take place in the system.

\subsection{The original simulation method}
\label{original}
To illustrate the algorithm, as in Ref.~\onlinecite{eletransf}, let us consider the reduction of cations, taking place according
to the electronic scheme of principle shown in Fig.~\ref{elelev} (a). The process is simulated by considering a spherical sample
representing an electrochemical half-cell (the cathode, in this case), consisting of $N_+$ cations and $N_-$ anions, with 
$N_+=N_-+Q_I$, where $Q_I$ is the 
net charge carried by the ions in the electrolyte. The whole sample is assumed to be neutral, because of the electronic surface 
charge density $\sigma_e=-Q_I/(4\pi R_e^2)$ uniformly distribute on the electrode. At first, the sample is equilibrated by 
GC-MC, excluding the electron transfer across the electrode/electrolyte interface. To conserve the net charge of the electrolyte
(and thus of the electrode), the GC-MC samples the addition and removal of neutral ion pairs. In a second stage, the electron 
transfer through the electrode/electrolyte interface is introduced. Following Ref.~\onlinecite{eletransf}, the electron transfer
involves only the cations in a narrow {\it active layer} of width $\delta R$ ($\delta R=d_{++}/2$ in the present case) on the 
electrolyte side next to the interface. The inherent kinetic rate $k_e$ is assumed to be the same for all active ions, and to 
vanish for any other ion. Moreover, $k_e$ is assumed to be independent of time. To solve the kinetic equation 
$dN_+=-k_e  N_+ dt$, each of the cations entering the {\it active layer} at time $t_0$ is assigned a lifetime $\Delta t$ 
extracted from the probability distribution:
\begin{equation}
p(\Delta t)=k_e \exp[-k_e \Delta t]
\label{pdel}
\end{equation}
Then, the GC-MC simulation is carried out with a fictitious {\it simulation time} $\tau$ increasing (uniformly) by one unit 
every $N=N_++N_-$ single-ion attempted MC displacements. Whenever the growing $\tau$ overtakes the $(t_0+\Delta t)$ time of any 
given active cation, that cation is removed from the active layer, and repositioned well inside the sample, with the new 
position selected from a probability distribution enforcing some additional condition, as discussed in 
Ref.~\onlinecite{eletransf}. In this procedure, removing the cation from the active layer represents its neutralisation by an 
electron crossing the electrode/electrolyte interface, while relocating this same cation inside the spherical sample corresponds
to the surfacing of a new cation from the bulk electrolyte, as required to preserve the overall sample neutrality.

During the simulation, active cations might spontaneously diffuse out of the active layer. In that case, they are removed from 
the list
of active cations, and the lifetime $\Delta t$ before relocation is no longer relevant. On the other hand, cations can enter the
active layer at any simulation time $\tau$. In such a case, they are assigned their own lifetime and are included in the list of
active particles. Slight variations of this algorithm give equivalent results, as verified by us in a few cases, provided they 
consist of the superposition of MC and the 
simulation of the kinetic equation Eq.~\ref{pdel}, with supplementary conditions that enforce the thermodynamic state of the 
electrolyte and the charge of the sample. The constant scale factor between MC and real time, calibrated once at the beginning 
of the simulation, allows to interpret the results in terms of real time units. Once again, more details and a short discussion 
are given in Ref.~\onlinecite{eletransf}.

We emphasise once again that at this stage, by our choice, the inherent electron transfer rate $k_e$: (i) is the same for all 
active ions; (ii) is independent of time and of the electrostatic conditions at the interface. Nevertheless, the current through
the interface shows a marked dependence electrostatic potential difference $\Delta V=\phi(R_e)-\phi(0)$, or, equivalently, on 
the electron charge density $\sigma_e$ on the electrode surface, as shown, for instance, in Fig.~6 of 
Ref.~\onlinecite{eletransf}. 

As a trivial matter of notation, we point out that the cathode half-cell will usually operate with an excess of cations on the 
electrolyte side, which corresponds to negative $\sigma_e$ and $\Delta V$. For this reason, we sometimes use $(-\sigma_e)$ and 
$(-\Delta V)$ as our variables, and even plots have been drawn with $(-\sigma_e)$ and $(-\Delta V)$ on their axes. The current 
density $j_r(R_e)$ is positive whenever the net transfer of electrons occurs from the electrode to the electrolyte cations.

The microscopic mechanism underlying the dependence of the current on the electrostatic potential drop across the half-cell is 
the progressive polarization of the electrostatic double-layer on the electrolyte side, which, with increasing charge density 
$-\sigma_e$ on the electrode surface, packs more cations in the active layer. The sharp rise of $j_r(R_e)$ with positive, 
moderate values of $-\sigma$ observed by simulation is reminiscent of but hardly exactly reproduced by basic electrochemistry 
relations like the 
Tafel\cite{tafe} and Butler-Volmer\cite{but, vol} equations. Besides the double-layer polarisation, in real systems the 
$j_r(R_e)$ versus $\Delta V$ dependence arises from multiple processes, one of which will be discussed in Sec.~\ref{fluctele}. 
The combination of these processes will also determine the overpotential value, as also discussed in the following.

\section{Simulation results and their interpretation}

Simulations have been carried out with the method briefly outlined in Sec.~\ref{original} and described in more detail in 
Ref.~\onlinecite{eletransf}. A series of half-cell samples have been considered, exploring a range of sample sizes, electrolyte
concentrations and surface charge densities overlapping those of the original publication. The main reason why computations have
been repeated is that grand-canonical Monte Carlo has been exclusively used in the present study, while  a combination of 
canonical and GC-MC simulations were carried out in the computations of Ref.~\onlinecite{eletransf}. The systematic usage of 
GC-MC, in which the number of particles fluctuates, is the reason why a round number of ions could not be imposed in any of the 
present simulations. The results discussed below concern samples of {\it about} $(4000+4000)$ ions at 2M concentration. For 
these particular cases, $8000$ attempted single particle MC steps (i.e., $1$ MC time unit) correspond to $0.28$ ns in real time
(see Tab.~IV in Ref.~\onlinecite{eletransf}), provided the amplitude $\Lambda$ of each single particle displacement is 
$\Lambda=10 d_{++}$ (in each Cartesian direction). Data for any other size and concentration are only briefly mentioned as a 
comparison.

\subsection{Accounting for the microscopic reversibility of chemical reactions}

For the sake of simplicity, Ref.~\onlinecite{eletransf} assumed that a single reaction takes place at each spherical interface, 
for instance the reduction of cations {M$^+$+e$^-$ $\rightarrow$ M} which defines the cathode. Because of this simplifying 
assumption, the reaction (reduction) rate $h_r$ at the cathode, once multiplied by (-e) for dimensional reasons, also represents
the current $j_r(R_e)$ flowing through the electrode/electrolyte interface. This coincidence may obscure the interpretation of 
the simulation results. For this reason, the
underlying one-reaction-only assumption will be somewhat relaxed, reflecting the fact that every M$^+$+e$^-$ $\rightarrow$ M 
reaction will always occur also in reverse: M $\rightarrow$ M$^+$+e$^-$. In general, the direct and reciprocal reaction will 
take place at different rates. The net current through the interface, therefore, will be determined by the difference 
between the direct and inverse reaction rates. It is important to emphasise that at this stage, the two complementary reactions 
concern the same M$^+$/M redox pair, taking place in the same half-cell. For this reason, while the rate of the direct reaction 
(which in our case is a reduction) is denoted by $h_r$, the rate of the inverse one will be denoted not with $h_o$, but with 
$h^{-1}_r$. In other terms, $h_r$ is the rate of electron transfer from the electrode to electrolyte cations, while $h^{-1}_r$ 
is the rate of electron transfer from neutral M atoms in solution to the metal electrode. According to the principle of 
microscopic reversibility, the equilibrium state correspond to the $h^{-1}_r=h_r$ condition. Deviations from this condition 
requires the application of an external voltage that goes under the name of {\it overpotential}. Every real electrochemical 
cell will also rely on a different reaction (X$^-$~$\rightleftharpoons$~X+~e$^-$, oxidation), taking place reversibly in the 
complementary half-cell. Consistently with the idea of focusing on half-cells, this complementary part of the process and 
device is not considered here.

In the present approach, the direct reaction, whose overall rate is $h_r$, will be simulated explicitly, the inverse one will be
accounted for implicitly, using plausible symmetry relations, as explained below. The major target of the simulation is the 
computation of the radial current density $j_r(R_e)$ across the electrode/electrolyte interface as a function of the 
electrostatic potential difference $\Delta V=\phi(R_e)-\phi(R=0)$ between the electrode surface and the centre of the spherical 
sample, which represents the bulk electrolyte. To this aim, one starts from the analysis of the reaction rate $h_r(\Delta V)$, 
whose determination is trivial, since the simulation program already counts the number of electron transfer events as a function
of MC and real time. As in Ref.~\onlinecite{eletransf}, the simulation algorithm is surface-charge-driven,\cite{linnea} but 
$\Delta V$ is determined at the same time of $h_r$ from the electrolyte charge density distribution, hence  the $h_r[\Delta V]$ 
relation is obtained point by point. 

\begin{figure}[!htb]
\vskip 0.7truecm
\includegraphics[scale=0.65,angle=-0]{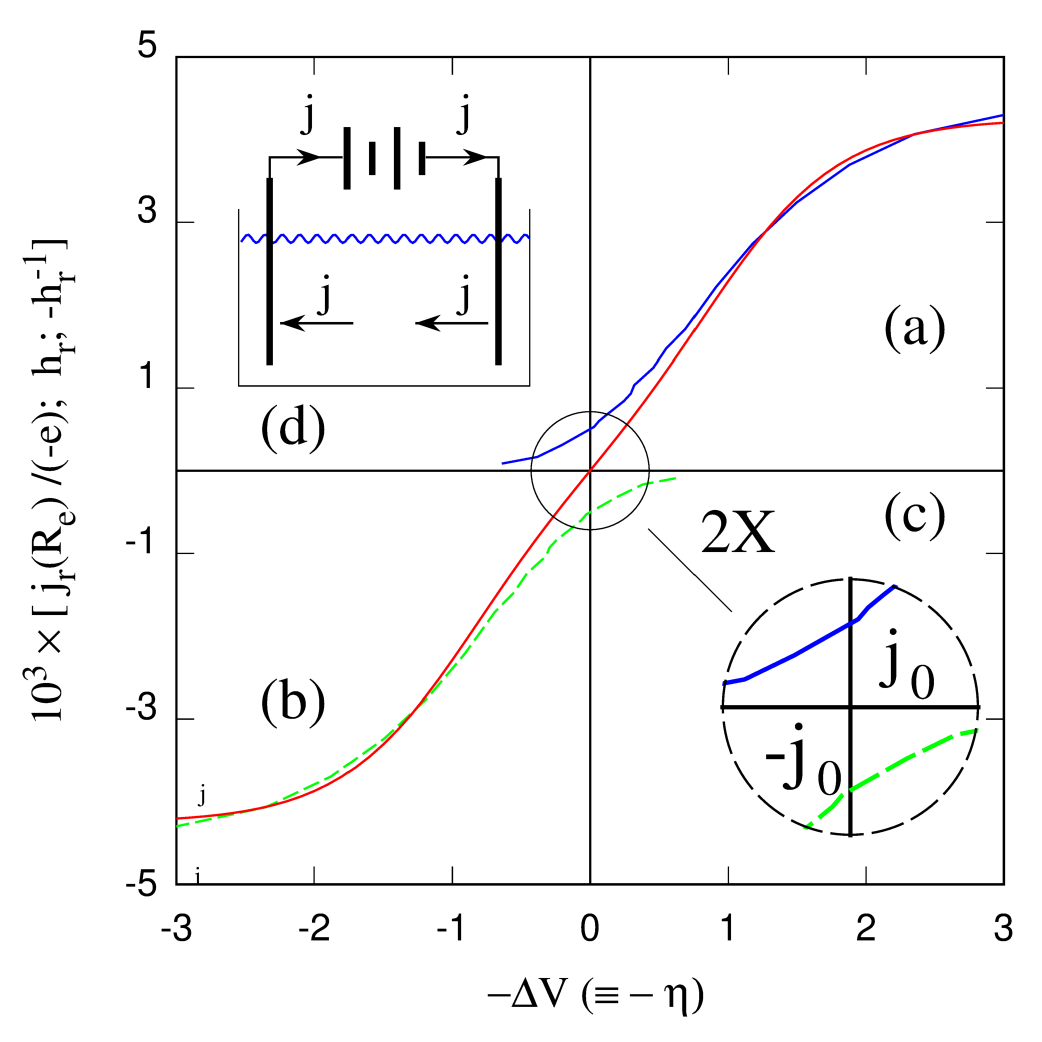}
\vskip 0.3truecm
\caption{Electron transfer rate through the surface of spherical metal electrodes of radius $R_e=21.78 \ d_{++}$ whose inside 
volume is filled by an electrolyte solution of 2M concentration. (a) blue line: overall rate $h_r$ of the cathodic 
{M$^++ e^- $ $\rightarrow$ M} reaction expressed in terms of events per unit area and unit time; (b) green dash line: overall 
rate $h_r^{-1}$ for the reciprocal reaction {M $\rightarrow$ M$^++e^-$}, also taking place at the cathode and displayed with a 
minus sign for clarity; (c) magnification of previous plots around the origin to illustrate the role of the exchange current 
$j_0$; 
(d) schematic drawing of the whole device of which the interface of interest (cathode) is one of two essential portions together
with the anode. The arrows show the direction of the electric current, which is opposite to the flow of electrons. Red line 
across the whole figure: $j_r(R_e)/(-e)=[h_r-h_r^{-1}]$ current density across the interface (in electrons per unit of time
and interfacial area).
}
\label{tafp}
\end{figure}

A typical result, referring the $\sim (4000+4000)$ ions at 2M concentration is shown in Fig.~\ref{tafp} (a). Starting from a low
$h_r$ value at moderately negative $(-\Delta V)$ (i.e., moderately positive $\Delta V$), the direct electron transfer rate $h_r$
increases monotonically with increasing 
$(-\Delta V)$, but the rapid increase shown at low positive $(-\Delta V)$ values, soon turns to a slower growth. This change of 
regime is forced by the viscosity (required to have a finite self-diffusion coefficient) which dominates the system dynamics at 
high flow, as briefly discussed in Ref.~\onlinecite{eletransf}.

Quantitative analysis also shows that at moderate (but not vanishingly small) $(-\Delta V)$ bias, the dependence of $h_r$ on 
$-\Delta V$ is virtually exponential (see below). In fact, $h_r$ does not vanish at any $\Delta V$ value, but remains positive 
while decreasing monotonically with $(-\Delta V) \rightarrow -\infty$. In particular, $h_r$ doesn't vanish either at the 
$\Delta V=0$ nor at the $\sigma_e=0$ point, 
because even at zero bias the set of active ions is not empty and the reaction will take place as long as $k_e > 0$. Notice 
that, for this symmetric electrolyte model, the $\sigma_e=0$ and $\Delta V=0$ will coincide for the ideally polarisable 
interface (i.e., $k_e=0$), but will differ for any $k_e >0$ value, since the non-vanishing flow of one type of charge will
make the profile of positive and negative charge to be inequivalent for any choice of $\Delta V$ and $\sigma_e$. Hence, the 
thermodynamically significant and experimentally measurable point of zero charge (pzc) coincides with $\Delta V=0$ only for the 
ideally polarisable interface ($k_e=0$), but, for the symmetric model under investigation, the difference is very small also at 
$k_e> 0$ and it will be neglected in what follows. In other terms, the $\Delta V=0$ and $\sigma_e=0$ points will be 
considered the same. In asymmetric systems, the pzc may differ even significantly from $\Delta V=0$, but the simulation 
approach will still be able to determine these two points virtually exactly, apart from the numerical error bar.

Again for the sake of simplicity, the original model of Ref.~\onlinecite{eletransf} did not specify the fate of the reduced M 
atom created by the neutralization of $M^+$ at the interface. This, in principle, makes it impossible to investigate by the same
method the time evolution and $\Delta V$ dependence of the inverse reaction, whose overall rate is $h^{-1}_r$. Nevertheless, it 
is possible to achieve an acceptable estimate of direct and inverse reaction rates for the $\Delta V$ close to the origin, which
is the only $\Delta V$ range for which the interplay of direct and inverse reaction matters. It is easy to realize that
virtually no direct reaction takes place at  $(-\Delta V )<<0$, since the active layer is virtually empty of cations, while at
$(-\Delta V) >>0$ every neutral M atom that forms in the active layer will be quickly pushed out by cations whose Coulombic
attraction to the electrode becomes overwhelming. It might be worth emphasising again that the oxidation reaction 
$X^-\rightarrow X+e^-$ is supposed to take place in the complementary half-cell, and the present discussion does not cover that 
part of 
any whole device.

At the pzc, i.e., $\Delta V = \phi(pzc)$, $M$ is present at non-vanishingly small concentration only at the interface, where it 
forms from the reduction of $M^+$ and it might survive a long time before undergoing the reciprocal reaction with the electrode,
or, much less likely, meeting a rare neutral $X$ atom to form again $M^+$ and $X^-$. Since each overall rate is virtually equal
to the product of the inherent rate times the number of active particles (apart from subtle many body effects), the following 
relation will be satisfied:
\begin{equation}
h_r/h^{-1}_r\propto n_a(M^+)/n_a(M)
\end{equation}
where $n_a(M)$ is the number of neutral atoms M in the active layer. At weak electrification of the interface, i.e., low 
$|\Delta V|$ and $|\sigma_e|$, the interaction of $M$ and $M^+$ with the electrode will be similar (apart from the difference in
size, which is felt at very short distance only). Then, any variation of $[\Delta V-\phi(pzc)]$ will affect the potential energy
of the charged species $M^+$ only, changing $n_a(M^+)/n_a(M)$ by a ratio $\exp{\{-\beta [\Delta V-\phi(pzc)]\} }$. Whatever is 
the sign of $\Delta V$ and $\phi(pzc)$, these conditions are satisfied if one assumes:
\begin{equation}
h^{-1}_r[(\Delta V-\phi(pzc))]=h_r[-(\Delta V-\phi(pzc))]
\label{symmetry}
\end{equation}
which summarises our simple theoretical model for the $\Delta V$ dependence of the reversible reaction 
{M$^+$+e$^-$ $\rightarrow$ M}. This relation is assumed to be valid on a limited range of $\Delta V$ around the origin. For the 
sake of simplicity, this range limitation has not been considered in drawing plots like Fig.~\ref{tafp}, as it is not explicitly
stated in equations written below like Eq.~\ref{inver}, but it has always been taken into account in applying 
Eq.~\ref{symmetry}. To be precise, a fit over the entire $\Delta V$ range explored by simulation is not needed, and a more 
restricted fit would be adequate. We followed this redundant route because the overall fit provides additional informations, 
which opens the way to investigate the reaction kinetics precisely where viscosity is important and the stationary state is 
determined by balancing the electron transfer rate and the diffusion of ions in the electrolyte.

In the present context, the symmetry relation Eq.~\ref{symmetry} can be used as follows. First, the $h_r$ dependence on 
$\Delta V$ in Fig.~\ref{tafp} (a) is fitted using the analytical expression:
\begin{equation}
h_r(\Delta V)= \frac{A\ \exp{[B (\Delta V-\phi(pzc))]}}{1+C\exp{[B (\Delta V-\phi(pzc))}]}
\label{direct}
\end{equation}
where $\phi(pzc)$ is the $\Delta V$ at the point of zero charge, which, in this case is virtually zero. The fit, using A, B and 
C as free parameters, is excellent, with a very low square deviation. It combines an exponential dependence of $h_r$ on
$(\Delta V-\phi(pzc))$ with the slowing down at the highest $j_r(R_e)$ or, equivalently, highest $-\sigma_e$. In the cases we
analysed, $C << 1$, saying that the onset of the viscosity-dominated regime is relatively slow. Moreover, $B$ turns out to be 
negative, consistently with the fact that $j_r$ grows with increasingly negative $\Delta V$.

Then, based on Eq.~\ref{symmetry}, one obtains:
\begin{equation}
h^{-1}_r(\Delta V)=  \frac{A \ \exp{[-B (\Delta V-\phi(pzc))]}}{1+C\exp{[-B (\Delta V-\phi(pzc))}]}
\label{inver}
\end{equation}
This expression is represented in Fig.~\ref{tafp}~(b) with a negative sign for clarity.

At any $(\Delta V-\phi(pzc))\approx \Delta V$ at the cathode, the net current through the interface (hence through the wire) is:
\begin{equation}
\frac{j_r(R_e)}{(-e)}=[h_r(\Delta V)-h^{-1}_r(\Delta V)]= \frac{2 A \ \sinh{[B(\Delta V-\phi(pzc))]}}
{1+C^2+2C\cosh{[B(\Delta V-\phi(pzc))]}}
\label{jofV}
\end{equation}
It is apparent that the $j_r$ current across each interface vanishes at $(\Delta V-\phi(pzc))=0$, not because the direct and 
inverse reactions stop, but because the currents they generate compensate each other. According to Eq.~\ref{direct} and 
Eq.~\ref{inver} the common absolute value of these opposite currents is $j_0=|A/(1+C)|$.

At $|B(\Delta V-\phi(pzc))|$ sufficiently low that the effect of viscosity is still negligible, $j_r(R_e)$ can be approximated 
as:
\begin{equation}
j_r(R_e)=j_0\left\{\exp{[B (\Delta V-\phi(pzc))]}-\exp{[-B (\Delta V-\phi(pzc))]}\right\}+o[(\Delta V-\phi(pzc))]^2
\label{approx}
\end{equation}
where $o[...]^2$ indicates the leading infinitesimal order of the difference between Eq.~\ref{jofV} and Eq.~\ref{approx}.
This expression is apparently the analog of Eq.~25.41 of Ref.~\onlinecite{atk}, which is referred to as the 
{\it Butler-Volmer equation}. If, at the same time, $[B (\Delta V-\phi(pzc))]$ is not too low, the first exponential 
significantly exceeds the second one (in other terms: $h^{-1}_r(\Delta V)<< h_r(\Delta V)$), hence:
\begin{equation}
j_r(R_e)\sim j_0 \exp{[B (\Delta V-\phi(pzc))]}
\label{eq10}
\end{equation}
which is equivalent to Tafel's equation:
\begin{equation}
\eta=-a \ log{\left(\frac{j_r}{j_0}\right)}
\label{taf}
\end{equation}
provided one identifies the overpotential $\eta$ with $(\Delta V-\phi(pzc))\approx \Delta V$, and recognises $j_0$ as the 
{\it exchange current} entering Tafel equation. In this context, the slope $a$ of Tafel equation is:
\begin{equation}
a=-\frac{1}{B}
\end{equation}
Then, simulation gives a way to compute $j_0$ and $a$ as a function of electrolyte composition and kinetic parameter $k_e$ by
fitting $h_r(\Delta V)$ over a limited overpotential range such that $h_r^{-1}$ is already vanishingly small, and viscosity is
still relatively unimportant. At the conditions 
of our simulations, the exchange current $j_0$ turns out to be small, and roughly proportional to the product of the kinetic 
coefficient $k_e$ times the number $n_a$ of cations in the active layer at $\Delta V=0$. For the electrolyte concentrations
covered by the present study, $n_a$ is low, and follows closely the trends displayed by the contact density of cations at the 
electrode surface,\cite{joel} which, in turn depends on the thermodynamic conditions, and of the pressure in particular, of the 
electrolyte far from the interface. 

Since, in the simulations discussed until now, $k_e$ is the same for all active ions and independent of time, the dependence
of $j_r(R_e)$ on $(\Delta V-\phi(pzc))$ described by Eq.~\ref{jofV} is entirely due to the polarisation of the ionic side of the
interfacial double-layer. In other terms, with increasing $-\sigma_e$, or, equivalently, increasing $-(\Delta V-\phi(pzc))$,
the number of active cations increases, and this is reflected by the parallel increase of $j_r(R_e)$. It might be worth pointing
out that the overpotential identified at this stage (polarization overpotential) is only one of the possible types of 
overpotential that are relevant for real systems. This important point will be commented upon again in the following 
Sec.~\ref{fluctele}.

\begin{figure}[!htb]
\vskip 0.7truecm
\includegraphics[scale=0.60,angle=-0]{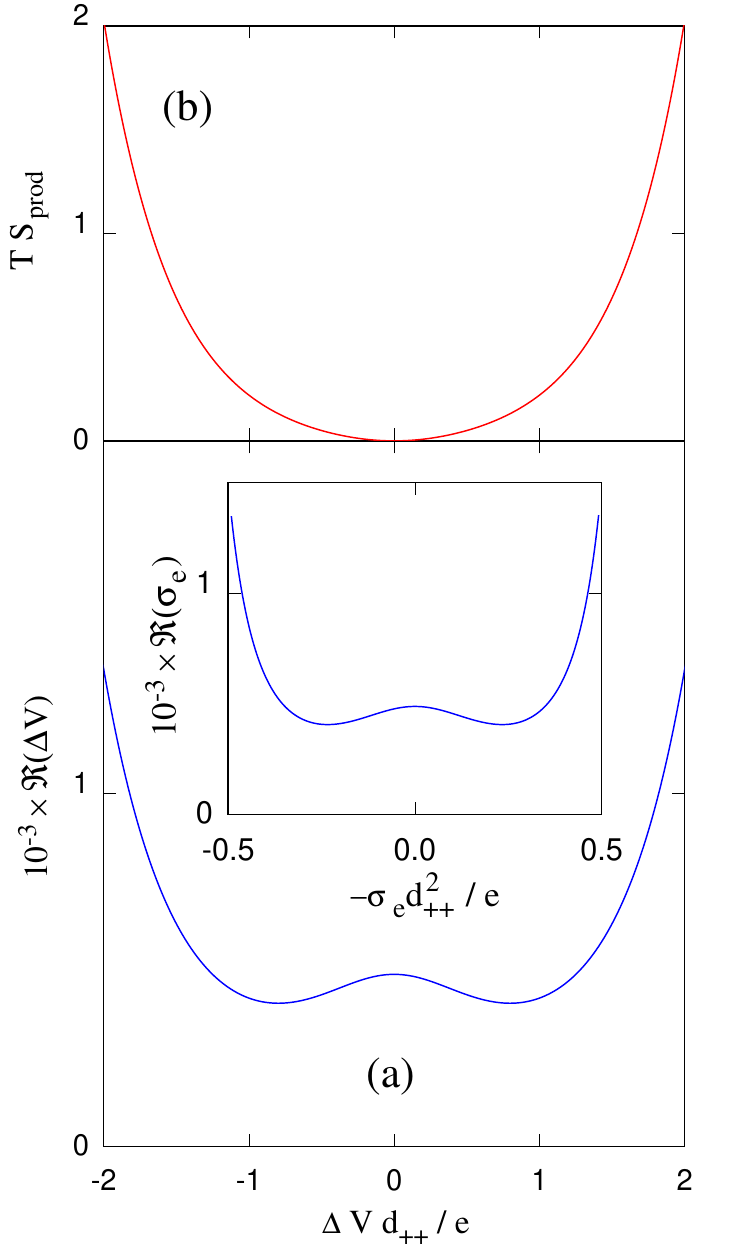}
\vskip 0.3truecm
\caption{
(a) Resistance per unit area ${\cal{R}}$ as a function of $\Delta V$ computed using Eq.~\ref{ohm} for the spherical half-cell 
at 2M electrolyte concentration and $k_e=0.01$. The inset in panel (a) gives the same ${\cal{R}}$ as a function of the 
interfacial charge density $-\sigma_e$. Panel (a) and its inset cover the same range of physical conditions, since 
$|\sigma_e|=0.5$ corresponds to $|\Delta V|=2.0$, in their respective units. (b) Rate of entropy production per unit surface
($S_{prod}$) for the same interface of panel (a), computed according to Eq.~\ref{produ} at the simulation temperature $T=298$ K.
The ${\cal{R}}$ data are given in the units listed in Sec.~\ref{mode}. The $TS_{prod}$ results are expressed in units of energy 
per unit area and unit time, see again Sec.~\ref{mode}.
}
\label{dohm}
\end{figure}

The data of $j_r(R_e)$ as a function of $\Delta V$ together with the analytical fit Eq.~\ref{jofV} allows us to express the DC 
resistance ${\cal{R}}$ of the interface as a function of $\Delta V$, using the differential form of Ohm's equation: 
\begin{equation}
{\cal{R}}[\Delta V]=\frac{d \Delta V}{d j_r(R_e)}=\frac{1}{\frac{d j_r(R_e)}{d \Delta V}}
\label{ohm}
\end{equation}
Here $j_r(R_e)$ and $\Delta V$ are meant to be the absolute values of these two quantities, and ${\cal{R}}$ is positive.
The results, again for the $N_++N_-\sim 8000$ ions is shown in Fig.~\ref{dohm} (a). With the present choice of units, 
${\cal{R}}$ turns out to be a relatively large number. This is due primarily to the fact that ${\cal{R}}$ refers to the 
microscopic interfacial area $d_{++}^2$. Using the rule for resistors connected in parallel, it is easy to see that, with
increasing interfacial area $w$, the resistance of the whole surface scales like $w^{-1}$.

An equivalent relation connecting ${\cal{R}}$ to the current density $j_r(R_e)$ can be determined implicitly through
Eq.~\ref{jofV}, using the fact that the dependence of $j_r(R_e)$ on $\Delta V$ is monotonic. We verified that the plot of
 ${\cal{R}}$ as a function of $j_r(R_e)$ (see the inset in Fig.~\ref{dohm} (a)) conveys  a message similar to the one given by 
${\cal{R}}$ as a function of $\Delta V$, which is shown in Fig.~\ref{dohm} (a). More precisely, the resistance ${\cal{R}}$ as a 
function of $\Delta V$ is nearly constant over the range for which the $j_r(R_e)$ versus $\Delta V$ dependence is nearly linear 
(see Figs.~\ref{tafp} and \ref{dohm}). 
Then, ${\cal{R}}$ increases rapidly at large  $|\Delta V|$ or, equivalently, large $j_r(R_e)$, where the system kinetics is 
dominated by viscosity. This behaviour emphasises the role of dissipation and irreversibility. Knowledge of the current density 
and of the Ohmic resistance of the interface allows to estimate the constant (in time) value of the heat dissipated across
the system. In a non-equilibrium, steady-state system, this heat dissipation corresponds to the entropy production rate 
per unit interfacial area of the half-cell:\cite{entrop}
\begin{equation}
S_{prod}=\frac{j_r(R_e)\Delta V}{T}
\label{produ}
\end{equation}
The entropy production rate, in particular, shows a parabolic range at moderate $j_r(R_e)$ and $\Delta V$, turning to a much 
faster growth at larger $\mid \Delta V\mid $ or $\mid j_r(R_e)\mid $ (see Fig.~\ref{dohm} (b)), where irreversibility is greatly
enhanced by the effect of viscosity.

\subsection{The electron transfer rate dependence on the electric field at the interface}
\label{fluctele}

The great appeal of simulation methods in soft-condensed matter physics relies on their ability to investigate correlations and
many-body effects bypassing the difficulties of traditional theoretical methods. In this respect, the versatility of the method 
of Ref.~\onlinecite{eletransf} is illustrated by extending the original approach to investigate subtle charge-charge 
correlations affecting the electric current flowing through the interface.

The simple model of Ref.~\onlinecite{eletransf} assumes a single inherent rate $k_e$, valid for every active cation of a given 
sample, irrespective of their distance from the electrode and from any other ion in the system. Then, the overall 
transfer rate will reflect the number of active ions, which, at constant T and concentration of the electrolyte, primarily 
depends on the average electrostatic potential at the interface. In reality, the inherent rate $k_e$ will also depend on a 
variety of parameters, including short range ion-ion correlations, structural and geometric features at the interface such as
steps, vacancies and impurities at the electrode surface, etc. One dependence that is necessarily present concerns the electric 
field acting on each active ion, which couples to the electrons jumping between the electrode and the electrolyte. Depending on 
its sign, the radial component of this field will enhance or reduce the barrier opposing the electron transfer. In the case of 
cations being reduced at the cathode, a positive  (i.e., pointing from the electrolyte towards the electrode) radial electric 
field will decrease the reaction barrier, enhancing the radial current density. The aim of this analysis is to show that the 
dependence of the inherent rates $k_e$ on the radial electric field is sufficient (and perhaps necessary) to introduce into the 
model essential features of real systems. We point out that this is the same current-voltage feedback mechanism assumed in 
widely known chemical physics textbooks\cite{atk} to justify the Butler-Volmer and Tafel equations. Although the textbook 
introduces the mechanism in terms of electrostatic potential, we prefer to express it in terms of radial electric field at the 
interface, and, instead of seeing it as the unique mechanism behind the current dependence on voltage (or, better, 
{\it overpotential}), we consider it as an additional mechanism besides the polarization of the electrolyte double-layer, that 
we already discussed in the previous Sec.~\ref{original}.

\begin{figure}[!htb]
\begin{minipage}[c]{\textwidth}
\vskip 0.7truecm
\begin{center}
\includegraphics[scale=0.50,angle=-0]{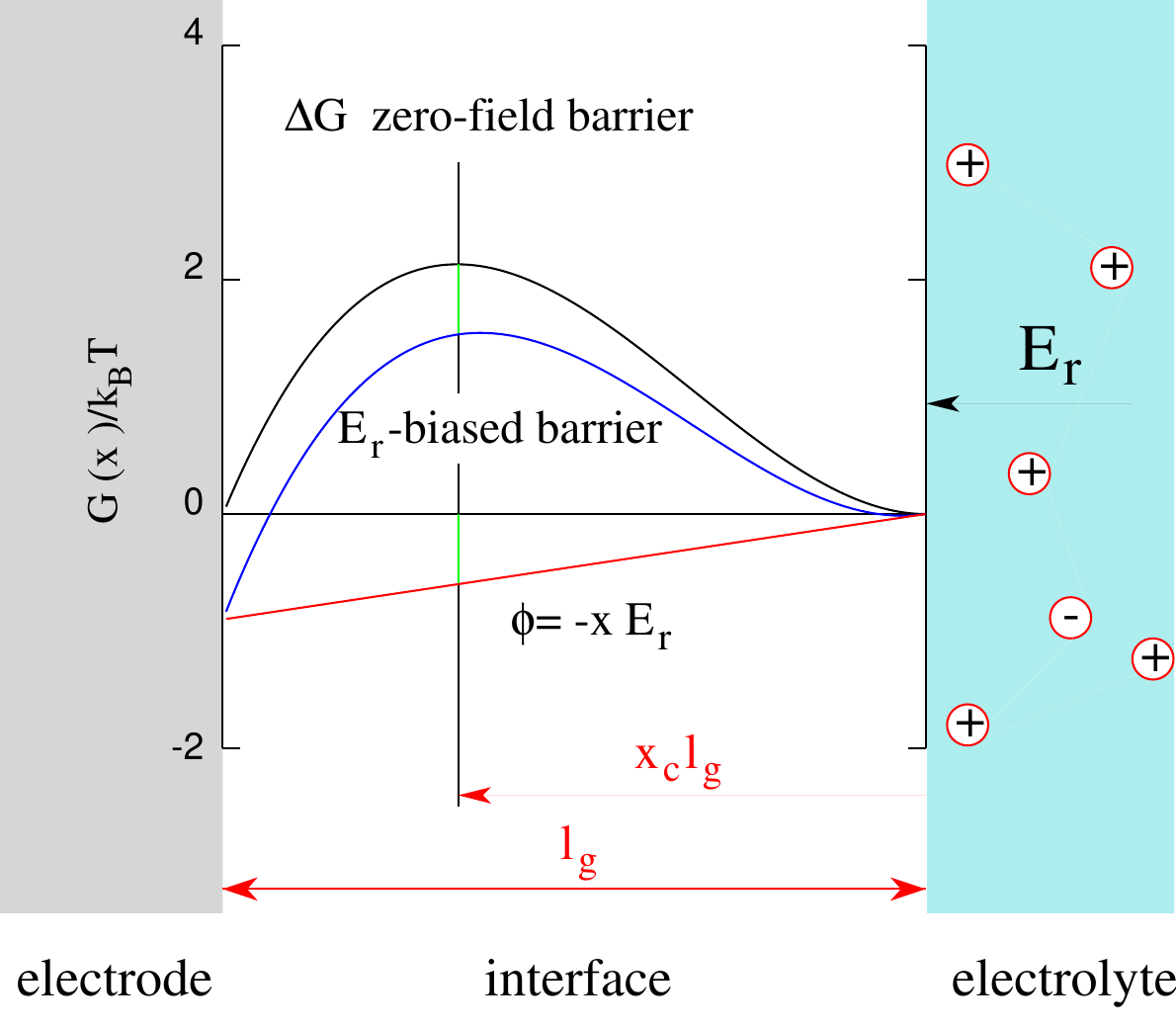}
\vskip 1.0truecm
\caption{Free energy profile along the constrained path of electrons moving from the electrode (left) to the electrolyte 
(right). Black line: unbiased profile, $E_r=0$. Blue line: profile biased by positive definite radial electric field ($E_r>0$). 
Red line: change of the electrons' free energy along the transfer path due to the radial electric field $E_r >0$. 
The interval $0\leq x \leq l_g$ represents the spatial gap that the electron has to cross to be transferred. This 
range between the electrode and the electrolyte is virtually empty and of nanometric length at most, hence the electric field is
considered constant over it. The two green segments on the vertical line at $x_c l_g=1.33\ d_{++}$ (corresponding to the 
coordinate of maximum free energy $\Delta G$) are of equal length.
}
\label{barrier}
\end{center}
\end{minipage}
\end{figure}

In what follows, we rely again on the spherical electrode set-up spanned by spherical coordinates whose origin is at the centre 
of the sphere. With this geometry in mind, the radial component $E_r$ of the electric field {\bf E} permeating any Coulombic 
system like the electrolyte, is orthogonal to the interface, and we will especially refer to its value at the position of ion 
{\it i} with $E_r(i)$. In this reference system, positive $E_r$ means that the 
electric field points outwards, and will pull electrons from the electrode to the active cations, enhancing the rate of cation
reduction. As long as the transition remains activated, the effect could be attributed to the lowering of the activation energy
$\Delta G$ by the electric field, that adds a linear variation of the potential energy profile along the reaction coordinate:
\begin{equation}
\frac{\delta \Delta G_i}{k_B T} =-\left[ \frac{e x_c l_g}{\epsilon k_B T} \right] E_r(i)\equiv -\alpha E_r(i)
\end{equation}
where $\delta \Delta G_i$ is the variation of the free energy barrier for the transfer of one electron to the cation whose label
is $i$; $l_g$ is the spatial gap separating metals and ions, and $x_c$ is the fractional position of the top of the barrier 
within this gap (see Fig.~\ref{barrier}), often referred to as the cathodic transfer coefficient. The parameter 
$\alpha=e x_c l_g / \epsilon k_B T >0$ is introduced to simplify the 
notation. Here we do not specify a detailed model for the barrier, but $\alpha$ will be used as a free parameter to gauge the 
strength of the barrier lowering. Then, the change in $k_e$ will depend exponentially on $\delta \Delta G / k_BT$, with negative
$\delta \Delta G/ k_B T$ corresponding to the enhancement of the electron transfer rate. If the transfer mechanism is 
primarily thermal, then the dependence will be expressed by the Arrhenius exponential factor 
$f_A=\exp{[-\delta \Delta G/k_BT]}$. If the transfer occurs by tunneling, then the exponential dependence will result from the 
WKB formalism.

Again for the sake of simplicity, here it is assumed that the electron transfer occurs by a thermal mechanism, and the 
dependence of $k_e$ on $E_r$ is given by:
\begin{equation}
\frac{k_e[E_r(i)]}{k_e^0}=\exp{\left[\frac{e x_c l_g E_r(i)}{\epsilon k_B T}\right]} 
\equiv \exp{\left [\alpha E_r(i) \right]},
\label{kei}
\end{equation}
where $k_e^0$ is the $k_e$ value at $E_r=0$, and $\alpha=e x_c l_g / \epsilon k_B T >0$. Typically $x_c \sim 0.5$ (i.e., the top
of the barrier is halfway between the 
electrode and the electrolyte) and $l_g$ is order of unit length $d_{++}$. Since at room temperature $k_B T$ is a small energy 
(again in our units) $\alpha$ can also be relatively large (a few units) without disrupting the basic picture of the electron 
transfer as an activated event. Notice that in the present model (as in the original one) the electrostatic potential and 
electric field are computed from the bare charges, while the dielectric constant is included in the parameter $\alpha$. To be 
precise, the most appropriate static dielectric constant to be used in this relation might not be the bulk water value 
($\epsilon=78.5$), but a significantly lower value suitable for the electrode-electrolyte interface. Here, however, the precise 
choice of this value is left to further studies, more focused on quantitative predictions than in the development of the 
simulation method. Another obvious but important observation is that the inherent transfer rate now is cation- and 
time-dependent, $k_e\equiv k_e[E_r(i|\tau)]$, since it will depend on the instantaneous radial electric field $E_r(i|\tau)$ 
acting on the active cation $i$ at the time of being reduced.

\begin{figure}[!htb]
\begin{minipage}[c]{\textwidth}
\vskip 0.7truecm
\begin{center}
\includegraphics[scale=0.65,angle=-0]{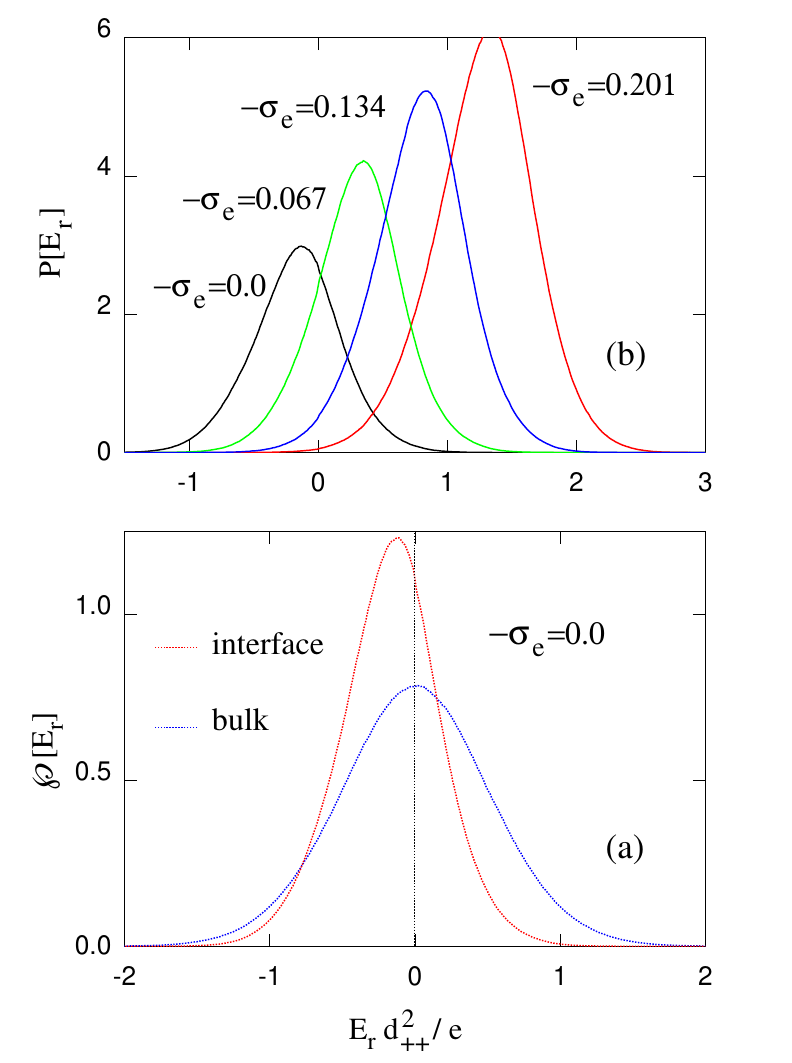}
\vskip 1.0truecm
\caption{(a) Probability distribution ${\cal{P}}[E_r]$ for the radial electric field $E_r$ on cations in fully polarisable 
samples ($k_e=0$). Blue line: computed on cations whose distance from the origin is $< R_e/2$, representing inactive bulk 
cations; red line: ${\cal{P}}[E_r]$ computed on active cations only, thus representative of the interface. Both curves computed 
on a sample with $\sigma_e=0$. The bulk distribution function doesn't depend on $\sigma_e$. (b) Probability distribution $P[E_r]$
on active cations as a function of the surface charge 
density $\sigma_e$. Negative values of $\sigma_e$ correspond to spherical electrolyte samples carrying an excess positive 
charge. Sample of 4000 (cation+anion) pairs at molar concentration 2M. The area under each curve is proportional to the number 
of active cations in the sample. Because of this non-standard normalisation, in panel (b) the distribution function is indicated
with $P[E_r]$ instead of ${\cal{P}}[E_r]$. Note the change of scale between (a) and (b).
}
\label{eledistr}
\end{center}
\end{minipage}
\end{figure}

The central quantity, therefore, is $E_r(i|\tau)$, i,e, the radial component of the electric field on the {\it active} cation 
{\it i} at simulation time $\tau$. Since this is a fluctuating variable, it will be characterised through its probability 
distribution function ${\cal{P}}(E_r)$. As shown in what follows, these probability distributions primarily reflect the 
ion-ion and ion-electrode correlation properties in proximity of the interface. As such, they will depend on temperature, bulk 
ion concentration and surface charge density $\sigma_e$.

\begin{table}[!htb]
\caption{Probability distribution ${\cal{P}}$ of the radial electric field $E_r$ on {\it active} cations. GC-MC simulations
on (4000+4000) ions samples, 2M concentration. $\langle E_r \rangle$: average; $\langle W \rangle$: standard deviation; 
$\langle Skew \rangle$: skewness; $\langle Kurts\rangle$: kurtosis of ${\cal{P}}(E_r)$. The average radial field at the
interface $E_r(Gauss)=-4 \pi \sigma_e$ from Gauss' law is given as a comparison. $\langle n_a \rangle$ refers to 
the whole sample.
}
\begin{center}
\begin{tabular}{|c|c|c|c|c|c|c|c|}
    \hline
 $-\sigma_e$         &0.00&        0.06712& 0.13425   & 0.20137& 0.26849     &0.3356  &0.40274\\
\hline
 $E_r(Gauss)$        &0.00&    0.8435     & 1.6870  &  2.5305  &  3.3740     &4.2175  &5.0610  \\
\hline
\multicolumn{8}{|c|}{k$_e=0$ \ \ \ \  }     \\
\hline
 $-\Delta V$              &-0.0010  & 0.2110  &  0.4359    & 0.6096  &  0.8308    &  1.0337 & 1.2658 \\
 $\langle n_a \rangle$   & 320     & 507     &   751      & 1037    &  1351      &  1676   & 2007   \\
 $\langle E_r \rangle$   &-0.1660  & 0.3016  &   0.7835   & 1.2721  &  1.7628    &  2.2601 & 2.7611 \\
 $\langle W \rangle$     &0.3493   & 0.3483  &   0.3641   & 0.3879  &  0.4185    &  0.4517 & 0.4873 \\
 $\langle Skew \rangle$  &-0.08131 &-0.0991  &  -0.1516   &-0.2059  & -0.2747    & -0.3601 &-0.4540 \\ 
 $\langle Kurts\rangle$  &3.3498   & 3.3317  &   3.3019   & 3.3088  &  3.2998    &  3.3061 & 3.3478 \\
\hline
\multicolumn{8}{|c|}{k$_e=0.01$ \ \ \ \  $\alpha=0$}     \\
\hline
 $-\Delta V$            & 0.0283   & 0.2452  & 0.4590     & 0.6758  & 0.9056     & 1.1727 & 1.4913 \\
 $\langle n_a \rangle$ & 317      & 504     &  741       & 1020    & 1323       &  1631  & 1930   \\
 $\langle E_r \rangle$ &-0.1646   & 0.3066  & 0.7914     & 1.2837  & 1.7834     & 2.2943 & 2.8198 \\
 $\langle W \rangle$   & 0.3490   & 0.3493  & 0.3626     & 0.3875  & 0.4178     & 0.4517 & 0.4883 \\
 $\langle Skew \rangle$&-0.0797   &-0.1106  &-0.1458     &-0.2021  &-0.2671     &-0.3436 &-0.4228 \\ 
 $\langle Kurts\rangle$& 3.3492   & 3.3428  & 3.3173     & 3.3095  & 3.2954     & 3.2888 & 3.2915 \\
\hline
\multicolumn{8}{|c|}{k$_e=0.01$ \ \ \ \  $\alpha=0.5$}     \\
\hline
 $-\Delta V$            &-0.0045   & 0.2303  & 0.4750     & 0.7261  & 1.0327     &  1.4688 & 2.1168 \\
 $\langle n_a \rangle$ & 316      &  501    &  736       & 1005    & 1282       &  1529   & 1679  \\
 $\langle E_r \rangle$ &-0.1664   & 0.3077  & 0.7937     & 1.2946  & 1.8134     &  2.3694 & 3.0093 \\
 $\langle W \rangle$   & 0.3504   & 0.3487  & 0.3634     & 0.3869  & 0.4168     &  0.4513 & 0.4875 \\
 $\langle Skew \rangle$&-0.0675   &-0.1120  &-0.1500     &-0.2011  &-0.2570     & -0.3073 &-0.3242 \\ 
 $\langle Kurts\rangle$&  3.3316  & 3.3442  & 3.3075     & 3.3089  & 3.2900     &  3.2476 & 3.1573 \\
\hline
\end{tabular}
\end{center}
\label{averages}
\end{table}

A baseline to gauge the role of $E_r(i)$ at any given electrochemical condition is provided by the corresponding 
distributions in the absence of electron transfer reactions, i.e., at ideally polarisable electrode/electrolyte interfaces. The 
simulation 
results for such a reference condition, carried out on samples of about (4000+4000) ions at 2M concentration are shown in 
Fig.~\ref{eledistr} (a) for four values of the electrode charge density $\sigma_e$. In this figure, the area under each $P(E_r)$
curve is equal to the average number of active ions per unit interfacial area $\langle n_a \rangle$. As before, the definition 
of {\it active ions} still relies on their distance from the interface being $\leq \delta R =d_{++}/2$, even though, at $k_e=0$,
no reaction takes place and, strictly speaking, all ions are inactive, even in the narrow active layer. As expected, 
$\langle n_a \rangle$ increases steadily with increasing $-\sigma_e$.

A quantitative characterisation of the distribution functions is summarised by a few parameters, such as the average radial 
electric field $\langle E_r \rangle$ (where $\langle ... \rangle$ indicates time and population average) and its standard 
deviation $W=\langle  E_r^2-\langle E_r\rangle^2\rangle^{1/2}$, as well as two standardised moments representing the skewness 
$Skew$ and kurtosis $Kurts$, defined as:
\begin{equation}
\langle E_r \rangle=\int_{-\infty}^{+\infty} E_r {\cal{P}}(E_r)d E_r
\end{equation}
\begin{equation}
W=\left[\int_{-\infty}^{+\infty} \left( E_r^2-\langle E_r \rangle^2\right) {\cal{P}}(E_r)d E_r\right]^{1/2}
\end{equation}
\begin{equation}
Skew=\frac{\int_{-\infty}^{+\infty} \left( E_r^3-\langle E_r \rangle^3\right) {\cal{P}}(E_r)d E_r}
{\left[\int_{-\infty}^{+\infty} \left( E_r^2-\langle E_r \rangle^2\right) {\cal{P}}(E_r)d E_r\right]^{3/2}}
\end{equation}
\begin{equation}
Kurts=\frac{\int_{-\infty}^{+\infty} \left( E_r^4-\langle E_r \rangle^4\right) {\cal{P}}(E_r)d E_r}
{\left[\int_{-\infty}^{+\infty} \left( E_r^2-\langle E_r \rangle^2\right) {\cal{P}}(E_r)d E_r\right]^{4/2}}
\end{equation}
where the normalised probability distribution ${\cal{P}}$ is given by:
\begin{equation}
{\cal{P}}(E_r)=\frac{P(E_r)}{\int_{-\infty}^{+\infty} P(E_r)d E_r}
\end{equation}
whose normalisation is $\int_{-\infty}^{+\infty} {\cal{P}}(E_r) d E_r=1$. The simulation results are collected in 
Tab.~\ref{averages}.

\begin{figure}[!htb]
\vskip 0.7truecm
\includegraphics[scale=0.50,angle=-0]{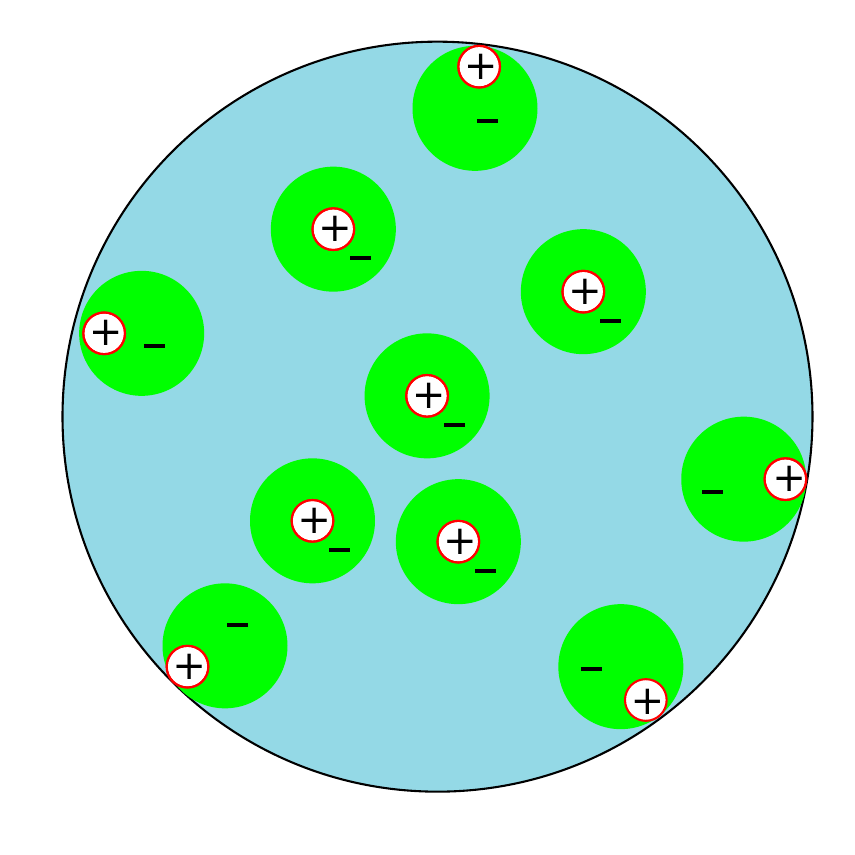}
\vskip 0.3truecm
\caption{Representative configuration of cations in the spherical sample. Each cation (small red circles with a {\bf +} inside) 
moves with its screening cloud (represented by a medium-size green circle) consisting of a fluctuating excess of anions whose 
density quickly decreases with increasing distance from the cation at their centre. The total charge of each screening cloud is 
equal to $-1$. When the cation is far from the interface, on average, the screening cloud is spherically symmetric with respect 
to the central cation.  Close to the interface, the spherical symmetry is broken and the relative position of the cation and its
screening cloud is such to pull the cation towards the electrolyte and away from the electrode.
}
\label{correla}
\end{figure}

In all cases, the ${\cal{P}}(E_r)$'s represent the distribution of electric field at the interface, since they are averaged on 
the active cations only. It is easy to verify by simulation that the electric field felt by ions close to the interface is 
predominantly radial, the other two components being small and vanishing on average. Moreover, the time average of the charge 
density of the electrolyte ions is spherically symmetric. In a static, mean-field picture, the average of the electric field at 
the interface $E_r(R_e)$ is given by Gauss' law as: $E_r(R_e)=-4\pi \sigma_e$, where $-4\pi R_e^2 \sigma_e$ is the net charge 
$Q_I$ of the spherical electrolyte. The ${\cal{P}}(E_r)$'s computed on active cations, however, are not $\delta$-functions at 
the value given by Gauss' theorem, but are 
broadened into a continuous distribution by thermal fluctuations and affected by correlations among the ionic positions.

Let us first comment on the results from the $k_e=0$ simulations. Remarkably, the $\langle E_r\rangle$ averaged on the ions
is systematically (and substantially) lower than the mean field value ${E}_r(R_e)=-4\pi \sigma_e$ averaged on the geometric 
surface. At the pzc, in particular, 
$\langle E_r\rangle < 0$.
The combination $[\langle E_r\rangle+4\pi\sigma_e]$ increases monotonically with increasing $-\sigma_e$ or, equivalently, 
increasing $-\Delta V$. The reason for this trend, expected but nevertheless interesting, is illustrated in Fig.~\ref{correla}. 
Both in the bulk and close to the interface, each ion gains Coulomb correlation energy by being surrounded by an excess of 
counter-ions whose distribution is described by a specific combination of pair distribution functions 
$\{g_{\alpha \beta}; \alpha, \beta= +, -\}$, whose volume integral is dictated by the perfect screening condition:\cite{joel}
\begin{equation}
\sum_{\beta} \rho_{\beta} Z_{\beta} g_{\alpha, \beta}({\bf r_{\beta}-r_{\alpha}}) d{\bf r_{\beta}}=-Z_{\alpha}
\end{equation}
valid for every ${\bf r_{\alpha}}$. In this equation, $\{ \rho_{\alpha}, \  \alpha=+,-\}$ is the number density of the two ion
species, and $\{Z_{\alpha},  \ \alpha=+,-\}$ is the charge of cations and anions, respectively. In other terms, each cation is 
surrounded by a cloud of excess anions whose net 
charge exactly screens the $Z_+$ charge of the ion under consideration. In a qualitatively pictorial way, this situation can 
be represented as in Fig.~\ref{correla}. While in the bulk the cloud of excess counter-ions $\beta$ is, on average, symmetric 
around each ions of type $\alpha$, it becomes asymmetric at the interface (Fig.~\ref{correle}). Although the picture is only 
qualitative, it is apparent that the cloud of excess counter-ions is positioned in a way to pull each cation away from the
interface, thus decreasing the average radial force $-4\pi \sigma_e$ pushing each cation towards the electrode surface. The 
difference between the simulation $\langle E_r\rangle$ and the Gauss value $-4\pi \rho_e$, which is the exact average of the
electric field over the spherical surface, is precisely the result of the asymmetric screening cloud at the interface, hence the
difference between $\langle E_r\rangle$ and $-4\pi \sigma_e$ is a manifestation of ion-ion correlations at the interface.

\begin{figure}[!htb]
\begin{minipage}[c]{\textwidth}
\vskip 0.7truecm
\hskip -2.0truecm
\begin{center}
\includegraphics[scale=0.60,angle=-0]{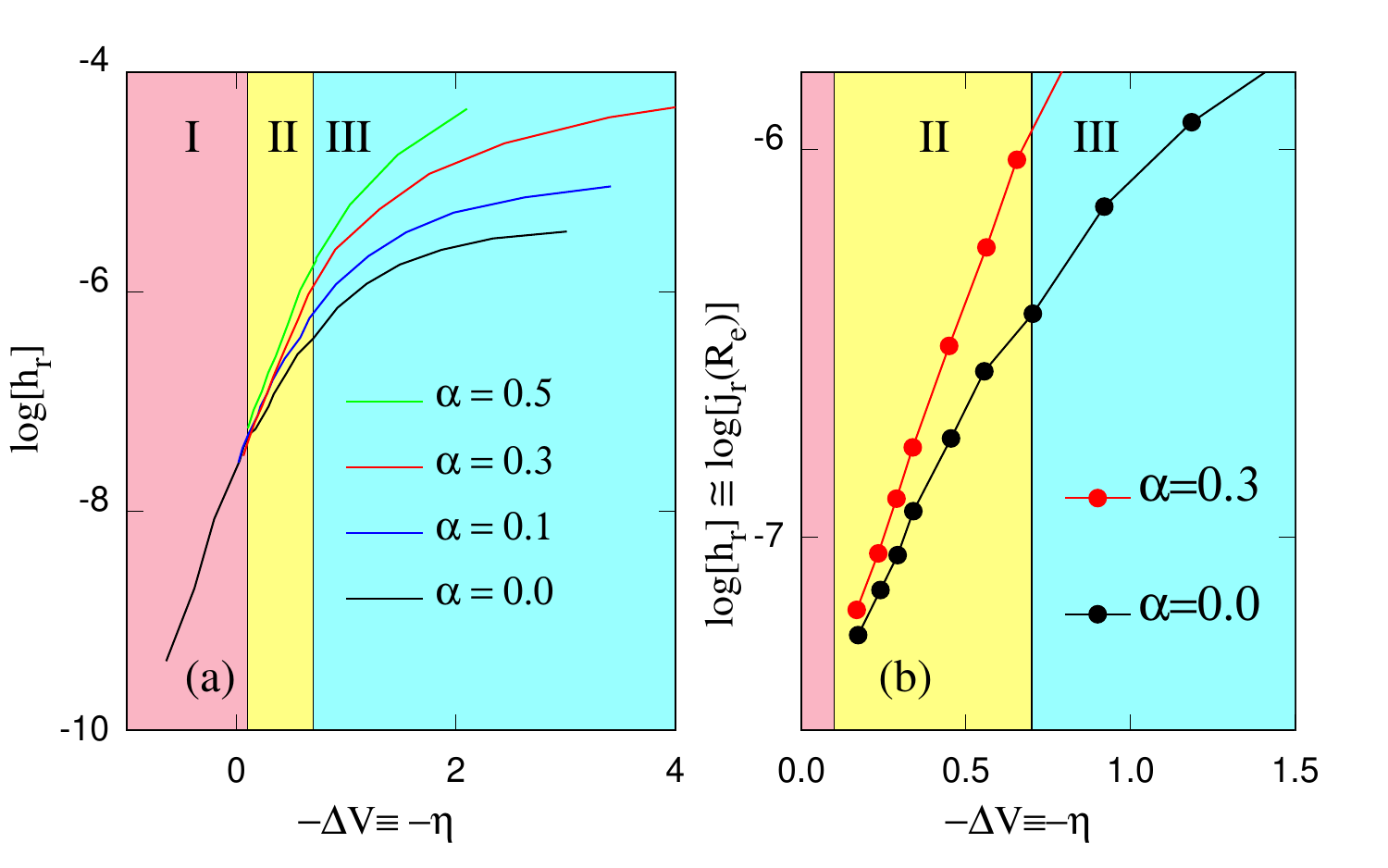}
\vskip 1.0truecm
\caption{Simulation results for the model with electric field feedback (Eq.~\ref{kei}) and four different values of $\alpha$.
Panel (a): data for the reduction rate (direct reaction) $h_r$ versus $\eta$ on a semilogarithmic scale. Panel (b): same 
plot on an expanded scale for the control case $\alpha=0$ and the $\alpha=0.3$ case. The three ranges I, II and III correspond
to three different regimes for $h_r$, $h_r^{-1}$ and viscosity, see text.
}
\label{overall}
\end{center}
\end{minipage}
\end{figure}

In a similar way, also the width of the probability distribution carries information on the ion-ion correlations. Strong
correlations limit fluctuations, thus fluctuations and correlations can be seen as antagonistic to each other. Then, the 
distribution of electric fields is significantly narrower at the interface than in the corresponding bulk electrolyte, as 
shown in Fig.~\ref{eledistr} (b), in which both probability distributions have been normalised in a way to sum to $1$. This
reflects the fact that interfaces strengthen the correlation among ions and make them longer range. Moreover, the scaled width 
$W/|\langle E_r \rangle|$ decreases rapidly with increasing $-\sigma_e$, since progressively charging the interface also 
increases correlation and reduces relative fluctuations.

It might be useful to point out at this stage that, both in the bulk and at interfaces, the distribution of fluctuating electric
fields acting on each ion also determines the tuning of the optical properties of the ions through the Stark effect.\cite{stk}
Optical spectroscopy and the analysis of the electron transfer rate at the interface, therefore, enjoy unexpected relations, as 
already  shown by a broad variety of experimental studies.\cite{song, cappel, san, roiati, ronca, johan, jacob} Therefore, the 
proposed simulation method could greatly help in analysing this relation at the atomistic level, thus enriching the data 
interpretation.

Let us turn to the simulation of partially unpolarisable interfaces (i.e., $k_e\neq 0$) using the rate equations in 
Eq.~\ref{kei}. At variance from the simplest model, the underlying transfer rate is now time dependent, and is different for 
each individual active cation. In what follows, the variable $k_e(i|t)$ will denote the instantaneous transfer rate at time 
$t$ for the cation whose label is $i$. Here $t$ is the real time (in ns), which is obtained by the MC time $\tau$ (measured in
attempted MC moves) by linear rescaling.

The kinetic equation to be solved, therefore, is:
\begin{equation}
\frac{d N_+(t)}{dt}= -\sum_{i \in active} k_e(i|t)
\label{kin}
\end{equation}
that replaces the more familiar:
\begin{equation}
\frac{d N_+(t)}{dt}= -k_e N_+(t)
\label{se}
\end{equation}
of the simplest model.

Because of the explicit time dependence of $k_e(i|t)$, the kinetic equation \ref{kin} cannot be solved as done for the simplest
model, but, at each MC step, every active ion will be tested for decay using its own $k_e(i|t)$. This different algorithm 
might limit the $\delta t$ of the time integration, which explicitly depends on the size of the MC attempted 
displacement (see Ref.~\onlinecite{eletransf}), slowing down the simulation. In the case of the present simulations, the 
computation is only marginally more time 
consuming than the original one for $\alpha=0$, and $\mu$s-long trajectories can still be simulated on a single core of a 
laptop. This statement, of course, is only true for the simple model of electrolyte that has been used, consisting of the 
primitive model with implicit solvent, with sample sizes up to $10^4$ cation-anion pairs. Larger samples, more complex 
electrolytes and especially explicit-solvent models might impose much stricter requirements on the efficiency of the 
computational algorithm or require much larger computer resources.

Provided the width $l_g$ of the interface is not negligible, and the barrier height is not overwhelming, the effect of
the radial electric field can be sizeable, especially if the surface charge density $-\sigma_e$ is not very small.  Geometric effects are important as well. As can be appreciated in Fig.~\ref{barrier}, the effect of the electric field is
strongest when the position of the barrier is closer to the electrode than to the electrolyte side. Some of these geometric 
aspects are qualitatively discussed in the textbook Ref.~\onlinecite{atk}.

Also in this case, explicit simulation have been carried out on sample of about $4000$ cations and $4000$ anions at 2M
electrolyte concentration, with $k_e^0=0.01$ and $\alpha=0,\ 0.1,\ 0.2,\ 0.3, \ 0.5$. For each system, the most relevant result 
is the relation between the radial current density $j_r(R_e)$ at the interface and the voltage $\Delta V$, expressing the 
electrostatic potential difference between the electrode and the centre of the electrolyte spherical cavity. Fig.~\ref{overall} 
(a) and (b) show, on a semi-logarithmic scale, the reaction rate $h_r$ which represents the raw quantity from which the net 
current $j_r(R_e)$ 
can be obtained using Eq.~\ref{symmetry}. The reason for this choice is that, contrary to $j_r(R_e)$, $h_r$ is positive 
over the whole $-\infty < \Delta V < \infty$ range and the results can be reported on a semi-logarithmic representation, whose 
importance will be apparent 
shortly. The whole figure covers a wide $\Delta V$ range, which has been divided into three parts. As apparent from 
Eqs.~\ref{symmetry}-\ref{approx}, $h_r$ is comparable to $h_r^{-1}$ in Range I, and the current density $j_r(R_e)$ has to be
worked out using Eq.~\ref{symmetry}. In Range II, Eq.~\ref{eq10} can be used to express $j_r(R_e)$ directly as
$j_r(R_e)/(-e)\sim h_r$. Range III is distinct from the previous two by the fact that in this range viscosity limits the ionic
conductivity, thus causing the deviation of the $j_r(R_e)$ dependence on $h_r$ from a simple exponential relation.

The results for $\alpha=0$, $k_e=0.01$ as a function of $\sigma_e$ represent the new baseline for assessing the quantitative 
and even qualitative effect of $\alpha > 0$. In general, these curves look largely as expected. The least predictable range is
in fact Range I, in the immediate vicinity of the pzc. Based on intuition, one could expect that the pzc is the condition at 
which the electric-field bias will switch from positive to negative, thus vanishing identically together with $j_r(R_e)$. 
However, since $\langle E_r \rangle$ is already negative at the pzc, the electric field bias will decrease the $h_r$ rate at 
the pzc slightly below its already small value. The validity of this statement is not apparent on the vertical scale of 
Fig.~\ref{overall}, but it is quantitatively confirmed by looking 
at tables for the reaction rates and conductivity data given by simulation. At $(-\sigma_e)>0$ and $(-\Delta V)> 0$, somewhat 
above the pzc (Range II) and below Range III, the electric field mechanism significantly enhances the charge transfer rate 
through the interface. The 
absolute, and to a lesser extent, also the relative enhancement grow with increasing $-\sigma_e$ and $-\Delta V$. As a result, 
in
Range II the electron transfer rate enhancement offsets the increasing effect of viscosity, extending the $\Delta V$ range over 
which the $h_r$ and $j_r(R_e)$ versus $\Delta V$ relation is virtually exponential, as can be seen in Fig.~\ref{overall} (b) 
which compares the $j_r(R_e)$ results for $\alpha=0$ and $\alpha=0.3$. This fact, in turn, stretches the $\Delta V$ range over 
which the Tafel and Butler-Volmer equations closely represent the computational data.

\begin{figure}[!htb]
\begin{minipage}[c]{\textwidth}
\vskip 0.7truecm
\begin{center}
\includegraphics[scale=0.70,angle=-0]{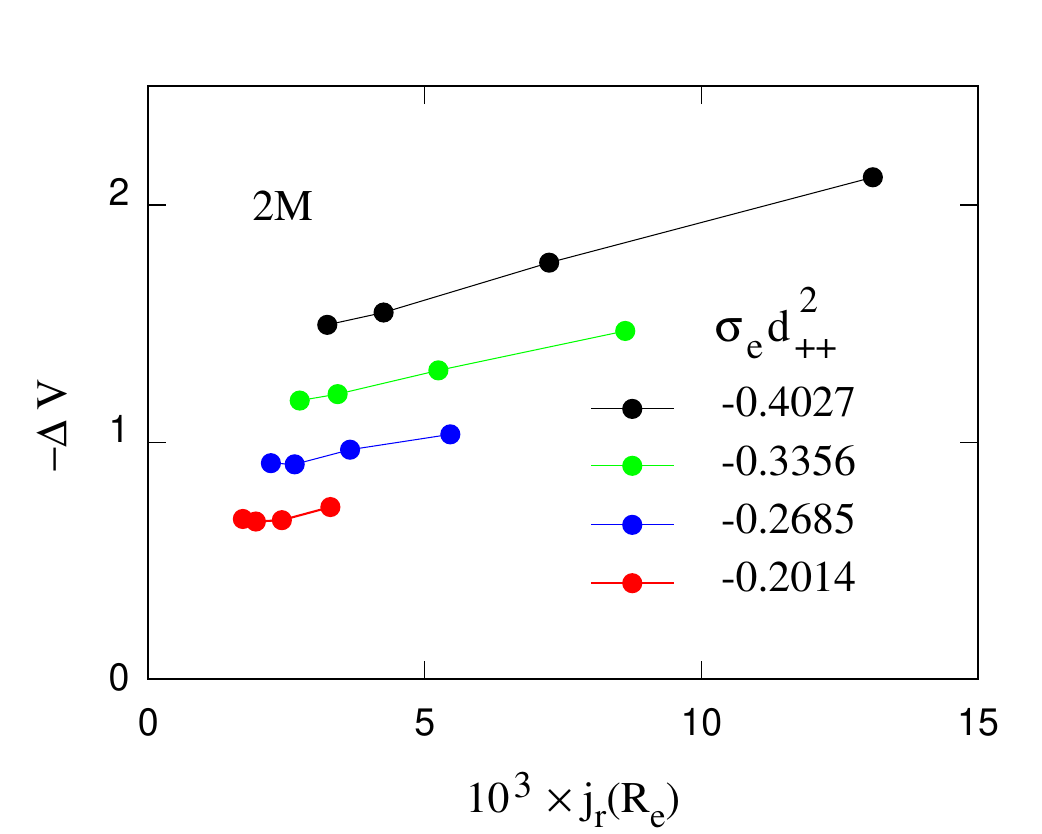}
\vskip 1.0truecm
\caption{Electrostatic potential difference $\Delta V=\phi(R_e)-\phi(0)$ as a function of the electric current density
$j_r(R_e)$ through the interface for different values of surface charge density $\sigma_e$ on the electrode. Samples of 
about (4000+4000) cations and anions at 2M electrolyte concentration.
}
\label{disrupt}
\end{center}
\end{minipage}
\end{figure}

As already mentioned, the efficiency of the algorithm might become more relevant for much larger or more complex systems. Hence,
it might be useful to explore options for replacing the variable $k_e(i|t)$ with an optimal single $\bar{k}_e$ value, allowing 
the usage of the algorithm of Ref.~\onlinecite{eletransf}. The large 
difference between $\langle E_r\rangle$ and $-4\pi \sigma_e$ implies that replacing all the $k_e(i|t)$'s in Eq.~\ref{kin} with 
the single value $\bar{k}_e=k_e^0\exp{(-4\pi \sigma_e \alpha)}$, would greatly exaggerate the enhancement (for positive 
$-\sigma_e$) or the slowing down (for negative $-\sigma_e$) of $h_r$ and $j_r(R_e)/(-e)$. As expected, a much better 
approximation is given by:
\begin{equation}
\frac{k_e(i|t)}{k_e^0}= \exp{\left [\alpha \langle E_r\rangle_0 \right]} 
\label{appx}
\end{equation}
which is independent of either $i$ and $t$. In this equation, $\langle E_r\rangle_0$ is the average of the radial electric 
field on active ions estimated for $\alpha=0$. This approach, of course, requires a preliminary simulation for $\alpha=0$, which
however can be relatively short because the simplest and most efficient algorithm (Eq.~\ref{se}) does apply, 
$\langle E_r\rangle_0$ converges quickly with simulation time, and the obtained $\langle E_r\rangle_0$ value can be used 
multiple times to 
simulate systems with different choices of $\alpha$. The difference with the results of simulations using the exact 
definition of $k_e(i|t)$ is quantitatively small, but nevertheless systematic. We suppose that this is due to the convexity of
the exponential function, but a formal proof is easily obtained only for a symmetric ${\cal{P}}(E_r)$ probability distribution.

\begin{figure}[!htb]
\begin{minipage}[c]{\textwidth}
\vskip 0.7truecm
\begin{center}
\includegraphics[scale=0.70,angle=-0]{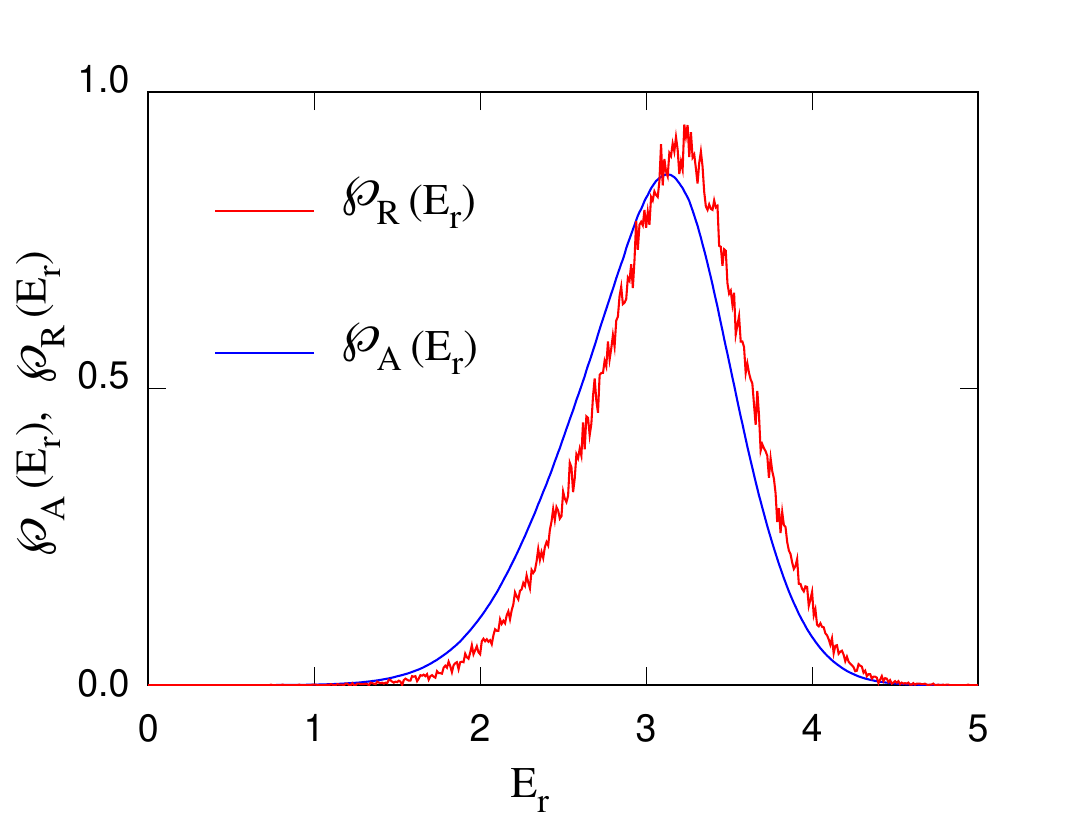}
\vskip 1.0truecm
\caption{Electron transfer simulation, (4000+4000) ion samples, 2M concentration, $k_e^0=0.01$, $\alpha=0.5$, $Q_I=2400$e.
Blue line: Probability distribution ${\cal{P}}_A[E_r]$ for the radial electric field $E_r$ on active cations.
Red line: Probability distribution ${\cal{P}}_R[E_r]$ for $E_r$ on active cations at the time of being reduced.
}
\label{correle}
\end{center}
\end{minipage}
\end{figure}

More subtle correlations connecting the cations' density distribution in space and the flow of charge through the interface with
fluctuations of the electric field can be seen among the results of Tab.~\ref{averages}. For instance, the flow of electrons 
through the interface depletes the population of active cations, reducing $\langle n_a \rangle$ with increasing $\alpha$. This, 
in turn, decreases the effectiveness of screening at the interface, and, at constant $-\sigma_e$, increasing $\alpha$ 
increases also the value of $(-\Delta V)$. As apparent from Tab.~\ref{averages}, this effect is far from negligible. In a 
similar way, the average $\langle E_r\rangle$ somewhat increases with increasing $\alpha$, again because the flow of charge 
decreases the effectiveness of screening at the interface. The fact that these trends are caused
by the steady flow of charge is confirmed by the observation that similar changes are observed by increasing $k_e$ at constant 
$\alpha=0$ as apparent by comparing the present results with those of Ref.~\onlinecite{eletransf}. The dependence of $\Delta V$ 
on the current density $j_r(R_e)$ is made explicit in the plot of Fig.~\ref{disrupt}. The simulation points align along nearly 
straight lines, that can be labelled with the value of the electronic charge density $\sigma_e$ residing at the 
electrode/electrolyte interface.

The dependence of $Skew$ and $Kurts$ on $\alpha$ is rather small, as shown again in Tab.~\ref{averages}. Since $Skew$ and 
$Kurts$ are already standardized moments, there is no need to further scale them with $\langle E_r(R_e)\rangle$, as instead it 
was done with the width $W$. Nevertheless, the trends in Tab.~\ref{averages} are
apparent, since the size of the changes, although small in absolute terms, are well beyond the error bars. In particular, $Skew$
increases and $Kurts$ decreases with increasing $\alpha$ (and thus with increasing interfacial current density), but no simple 
interpretation of these observations is available yet.

A further correlation can be highlighted by comparing the probability distribution ${\cal{P}}_A(E_r)$ computed over all $n_A$ 
active cations, and ${\cal{P}}_R(E_r)$, restricted to the $n_R$ cations at the time of their reduction. As shown in 
Fig.~\ref{correle}, the second curve (red line) is shifted to the right with respect to the first one (blue line), confirming 
that, as expected, the M$^+$ cations that, by fluctuation, experience a stronger electric field, are more likely to be reduced 
to M. This is possibly the most direct representation of how and how much the fate of active cations is affected by fluctuations
in the radial electric field at the interface. Starting from the definition of ${\cal{P}}_A$ and ${\cal{P}}_R$, one can derive:
\begin{equation}
\frac{n_R {\cal{P}}_R}{n_A {\cal{P}}_A}=k_e^0 \exp{[\alpha E_r]} 
\end{equation}
which provides a consistency relation to check the simulation results. The results presented in this section fully comply with 
this condition.

Although most of the results of this subsection could be expected, the analyses of these last paragraphs illustrate how deep
and detailed can be the investigation by the present method of the interplay among properties such as ionic correlations, 
fluctuations of the electric field at the interface, disruption of screening due to the flow of charge, as well as stochastic 
aspects of the electron transfer through the interface arising from quantum mechanical processes.

\section{Summary and conclusions}
Electrochemistry, together with its numerous applications, is a paradigmatic example of a multidisciplinary subject, firmly 
rooted in physics and chemistry, encompassing quantum and classical regimes, stretching over several size and time scales, and 
concerning systems that in many (if not most) cases are aptly considered as complex. A recently proposed Monte Carlo simulation 
method holds the promise of a balanced account of several of these aspects, thus reflecting properties and behaviours of real 
electrochemical systems more directly and faithfully than traditional simulation methods, including, in some instances, even 
ab-initio MD. The original methods has been able to deal with relatively high numbers of ions (up to 16,000 per half-cell) 
through its usage of an implicit solvent model. More importantly, it easily attained multi-micro-second time scales thanks to 
its kinetic interpretation of the underlying MC algorithm. Admittedly, the first formulation also adopted many simplifying 
assumptions. Nevertheless, it has been convincingly shown that the proposed model of electrochemical half-cell, consisting of a 
single electrode/electrolyte interface reaches a robust steady-state, in which charge flows across the interface at nearly
constant current density $j_r(R_e)$, with a realistic matching of electronic and ionic conductivity.

The present study delve into subtle aspects of the model and of the interpretation of the simulation results. For the sake of
definiteness, also in this case the model, the simulations and the analysis of the data are focused on the case of cations
being reduces by electrons pulled from the electrode (cathode). First of all, the present study provides a better definition of 
the model, showing that, through the incorporation of microscopic reversibility, concepts such as the overpotential, overall 
irreversibility, and entropy production arise naturally from the method. In the present study, in particular, the microscopic
reversibility is introduced implicitly, based on an antisymmetry assumption expressed by Eq.~\ref{symmetry}. In this respect,
more than introducing a new approach, the present study proposes a new way of interpreting the raw simulation data, consisting 
of the $h_r$ curve as a function of the electrostatic potential parameter $\Delta V=\phi(R_e)-\phi(0)$. The analysis of the rate
for the direct (M$^+$+e$^-$ $\rightarrow$ M) and inverse (M $\rightarrow$ M$^+$+e$^-$) reactions at the interface allow to 
identify crucial quantities such as the exchange current $j_0$ and the overpotential $\eta$ which, for this simple, symmetric 
model of electrolyte, virtually coincides with $\Delta V$.  The process of identifying $j_0$ and $\eta$ from the analysis of 
currents and electrostatic properties close to the origin of the $[\eta; j_r(R_e)]$ plane provides a novel view of the origin 
of Tafel and Butler-Volmer equations, independent of any empirical input and only based on the numerical analysis of the 
simulation data. Moreover, further analysis of the $j_r(R_e)$ versus $\Delta V$ relation allows to determine the electrical
resistance per unit area ${\cal{R}}$ associated to the flow of charge across the interfacial region of the half-cell. This, in 
turn, provides information on irreversibility and entropy production due to the flow of a steady electric current through the
half-cell.

The second part of the study, presented in Sec.~\ref{fluctele}, aims at showing how new features can be incorporated into the 
model, enhancing its realism and thus extending its reach. The electron transfer process, in particular, involves a free energy 
activation. Hence, at constant temperature, its rate depends on the free energy profile along a suitable reaction coordinate 
describing the motion of one electron from/to the metal to/from an ion next to the electrode surface. This dependence is 
expected to follow the well known Arrhenius law.\cite{atk} In this respect, fluctuating electric fields at the 
electrode/electrolyte interface play an important role, affecting the transition barrier as schematically indicated in 
Fig.~\ref{barrier}. Under the conditions described in the text, the fluctuating electric fields favour the electron 
transfer (see Fig.~\ref{overall} (a)), and the rate enhancement becomes more sizeable at high values of 
the current density $j_r(R_e)$ through the interface, where viscosity tends to slow down this same electron transfer rate. 
Semi-logarithmic plots of the current density $j_r$ versus overpotential $\eta$  (see Fig.~\ref{overall} (b)) show that the 
partial compensation of these two 
effects, i. e. the transfer-rate enhancement by the electric field and the slowing down of the ionic current by viscosity, 
results in a wider overpotential range in which $j_r$ versus $\eta$ is approximately linear with respect to the simpler model 
in which the fluctuating electric field at the interface plays no role, and only the double-layer polarisation affects the 
size of the electric current.

It is important to remark that, although presented in somewhat different terms, the mechanism introduced in Sec.~\ref{fluctele} 
to account for the effect of electric fields on the electron transfer rate is the same considered in popular textbooks like 
Ref.~\onlinecite{atk} to provide a plausible, semi-heuristic derivation of the Butler-Volmer and Tafel equations. Simulation
treats the two mechanisms, i.e., the double-layer polarisation and the barrier lowering mechanism, on the same footing, and
allows to assess their relative role in defining the value of the combined overpotential.

In this context, a further observation is suggested by the comparison of the different curves in Fig.~\ref{overall} (a) and (b).
As 
expected, with increasing $\alpha$, the same $j_r(R_e)$ current density is obtained at a lower $|\eta|$ value. More in general, 
the combination of different mechanisms that favour conductivity (in this case, the double-layer polarisation and the electric 
field feedback) decreases the size (i.e., the modulus) of the overpotential, and decreases the dissipation associated to the 
flow of current. Simulation allows to switch-on / switch-off the individual mechanisms that affect the electron transfer rate
and the ionic conductivity, and could provide useful insight on how the different mechanisms combine to determine the properties
of the half-cell, including conductivity, overpotential and dissipation.

The quantitative analysis of the effect of fluctuating electric fields and of their coupling with the electron transfer
mechanism highlights new aspects related to ion-ion and ion-interface correlations. First of all, the radial electric field 
averaged over all active ions is significantly less than the average over the whole spherical interface, as computed from the 
electrode charge density $\sigma_e$ according to Gauss theorem ($E_r(R_e)=-4\pi \sigma_e$), and this systematic difference can
be unambiguously attributed to the effect of the interface on the ion-ion correlations. Then, the correlation between ions and 
the electrode manifests itself in several additional ways, two examples being: (i) the probability distribution ${\cal{P}}(E_r)$of the radial field $E_r$ felt by active cations is significantly narrower than the same distribution computed for ions in the 
bulk electrolyte, pointing to the strong enhancement of correlations due to the presence of the sharp interface; (ii) the 
average of the radial electric field computed on M$^+$ cations at the time of their reduction to neutral M atoms is higher than 
the corresponding average computed on all active ions (see Fig.~\ref{correle}). Since the model electrode is a metal conductor, 
peaks in the radial electric field at the interface point to correspondingly large fluctuations in electronic surface charge of 
the electrode, thus highlighting a positive correlation between the instantaneous fluctuations of the electrode charge density 
and the electron transfer rate. A further remark is that fluctuating electric fields on the electrolyte side can be probed 
through the Stark tuning of the optical transitions involving the electronic levels of the electrolyte ions. This, in turn, 
establishes a further link between microscopic properties of the electrons and especially electrolyte ions, and quantities that 
can be measured by experiments. 

An important result coming both from the first and second part of the investigation is the evidence of how the flow of charge 
disrupts the Coulombic screening at the interface, as apparent from the dependence of $\Delta V$ as a function of $j_r$ at 
constant $\sigma_e$ shown in Fig.~\ref{disrupt}.

A few additional remarks that concerns the whole study and its results are as follows.

The improvements in the analysis and interpretation of the simulation results, together with the introduction of further aspect 
towards a less idealised and more realistic model, will progressively reduce the gap separating approaches to simulate
electrochemical systems from models of electron transfer in purely electronic systems (see Ref.~\onlinecite{jaco}), and might 
enhance the role of simulation in the design and analysis of electrochemical devices.

It is likely that already in the near future machine-learned approaches\cite{bodo} will transform this electrochemical branch of
simulation as they have done in several other simulation sub-fields. Nevertheless, the present kinetic model might provide a 
better basis for the incorporation of machine learning capabilities than more traditional, classical or ab-initio MD methods.

Once again, we point out that our simulations are carried out under surface-charge-control, while experiments are
generally carried out under potential-control. We admit that our choice has been motivated primarily by convenience, since
controlling the charge (and hence the average force) at the electrode surface is the simplest option to implement. The two 
choices are not strictly the same, since using one or the other is equivalent to using different ensembles in estimating the
thermodynamic properties (volume, energy, pressure, T) of a fluid system. Using two different ensembles does not 
change the relation between primary thermodynamic functions, while their fluctuations and response functions are affected.
In this way, the primary target of our simulations, i.e., the relation between the current charge density $j_r(R_e)$ and
the electrostatic potential (or overpotential) $\Delta V=\phi(R_e)-\phi(0)$ is correctly estimated, while fluctuations and 
response functions, which also are of interest, do depend on the choice between the two control modes. We are confident that our
approach could be adapted to potential-control conditions, since suitable algorithms are available and already used both in 
classical simulation\cite{scalfi} and even in ab-initio MD\cite{lozovoi, pasquarello} of electrified interfaces.

\begin{acknowledgments}
D.V.-D. and  N.C.F.-M. acknowledge the Deutsche Forschungsgemeinschaft (DFG, German Research Foundation) for funding through the research group FOR 2982–UNODE, Project number 413163866. N.C.F.-M. gratefully acknowledges the ECHELON Project from the Carl Zeiss Foundation for financial support. R.C.-H. acknowledges the European Union for funding through the Twinning project FORGREENSOFT (grant no. 101078989 under HORIZON-WIDERA-2021-ACCESS-03). R.C.-H. acknowledges funding from SFB-TRR146 of the German Research Foundation (DFG)–Project No. 233630050. R.C.-H and N.C.F.-M. thank Kurt Kremer for insightful comments.
\end{acknowledgments}

\end{document}